\documentclass[A4paper]{JHEP3}
\usepackage{epsfig,array,cite,latexsym,amsmath,enumerate}
\preprint{INT-PUB 02-40, UW/PT 02-14}
\newcommand{\beq}{\begin{equation}}
\newcommand{\eeq}{\end{equation}}
\newcommand{\beqa}{\begin{eqnarray}}
\newcommand{\eeqa}{\end{eqnarray}}
\newcommand\eqn[1]{\label{eq:#1}} 
\newcommand\Eq[1]{Eq.~(\ref{eq:#1})} 
\newcommand\half{\textstyle{\frac{1}{2}}}

\newcommand\bra[1]{\langle #1 \vert}
\newcommand\ket[1]{\vert #1 \rangle}
\newcommand\braket[2]{\langle #1 \vert #2 \rangle}
\newcommand{\digc}{\kern-1.5pt \raisebox{1.02ex}{$\cdot$}  \kern1.5pt
  \raisebox{0ex}{${\mathbf\cdot}$}\kern1.5pt \raisebox{-1.02ex}{$\cdot$}} 
\newcommand{\CD}{{\cal D}}
\newcommand{\CO}{{\cal O}}
\newcommand{\CL}{{\cal L}}
\newcommand{\CR}{{\cal R}}
\newcommand{\bfn}{{\bf n}}
\newcommand{\bfr}{{\bf r}}

\newcommand{\bfm}{{\bf m}}

\newcommand{\xh}{{\bf \hat x}}
\newcommand{\yh}{{\bf \hat y}}
\newcommand{\zh}{{\bf \hat z\,}}
\newcommand{\Tr}{\mathop{\rm Tr}}
\newcommand{\sla}[1]%
        {\kern .25em\raise.18ex\hbox{$/$}\kern-.75em #1}
\newcommand{\mybar}[1]%
        {\kern 0.8pt\overline{\kern -0.8pt#1\kern -0.8pt}\kern 0.8pt}

\newcommand{\drawsquare}[2]{\hbox{%
\rule{#2pt}{#1pt}\hskip-#2pt
\rule{#1pt}{#2pt}\hskip-#1pt
\rule[#1pt]{#1pt}{#2pt}}\rule[#1pt]{#2pt}{#2pt}\hskip-#2pt
\rule{#2pt}{#1pt}}
\newcommand{\Yfund}{\raisebox{-.5pt}{\drawsquare{6.5}{0.4}}}
\newcommand{\Ybarfund}{\mybar{\raisebox{-.5pt}{\drawsquare{6.5}{0.4}}}}%
\newcommand\fverb{\setbox\pippobox=\hbox\bgroup\verb}
\newcommand\fverbdo{\egroup\medskip\noindent%
                        \fbox{\unhbox\pippobox}\ }
\newcommand\fverbit{\egroup\item[\fbox{\unhbox\pippobox}]}
\newbox\pippobox

\title{ Supersymmetry on a Spatial Lattice}

\author{David B. Kaplan \\ Institute for Nuclear Theory, University of Washington,
  Seattle, WA 98195-1550 \\Email: \email{dbkaplan@phys.washington.edu}}

\author{Emanuel Katz and Mithat \"Unsal\\ Dept. of Physics, University
  of Washington, Seattle, WA 
  98195-1560 \\ Email: \email{amikatz@phys.washington.edu},
  \email{mithat@u.washington.edu}}

\abstract{
We construct  a  variety
of supersymmetric gauge
theories on a spatial lattice, including $N=4$ supersymmetric
Yang-Mills theory in 3+1 dimensions.  Exact lattice
supersymmetry greatly reduces or eliminates the need for fine tuning
to arrive at the desired continuum limit in these examples.  }

\begin{document} 

\section{Searching for accidental supersymmetry}
\label{sec:1}

There are numerous fascinating features in strongly coupled
supersymmetry, supergravity and string theory that would benefit from
numerical investigation. Furthermore, it is desirable to know whether
a well defined nonperturbative description of these theories
exists. For these reasons much effort has been devoted to the 
formulation of lattice versions of supersymmetric field theories  
\cite{Dondi:1977tx,Banks:1982ut,Elitzur:1982vh,Bartels:1983wm,Golterman:1989ta,Huet:1996pw,Montvay:1996ea,Nishimura:1997vg,Neuberger:1998bg,Nishimura:1998hu,Bietenholz:1998qq,Kaplan:1999jn,Fleming:2000fa,Fleming:2000hf,Farchioni:2001wx,Catterall:2001wx,Catterall:2001fr,Fujikawa:2002ic}.
To date, this work has led to limited practical success, confined
primarily to some $1+1$ dimensional theories, or $N=1$ supersymmetric
Yang-Mills (SYM) theory in $3+1$ dimensions.

The origin of the problem is that supersymmetry is part of the
super-Poincar\'e group,  which is explicitly broken by the lattice. 
Ordinary Poincar\'e invariance is also broken by the
lattice, but  due to the crystal symmetry of the lattice, relevant operators which could spoil 
the emergence of Poincar\'e symmetry in the continuum limit are
forbidden. In a supersymmetric theory, however, the lattice point
group is never sufficient  to forbid relevant supersymmetry violating operators.

One would like to implement
exact symmetries on the lattice that ensure that supersymmetry emerges as
an ``accidental''  continuum symmetry of the theory.  In four
dimensions, accidental supersymmetry can be achieved
for  $N=1$ supersymmetric Yang-Mills (SYM) theory, a theory without
spin zero bosons. (See ref. \cite{Montvay:2001aj} for discussion and
references.)  This is because for a generic $SU(k)$ gauge theory with an
adjoint Weyl fermion,
the only relevant operator which violates $N=1$
supersymmetry is a fermion mass term, which
can be forbidden by a discrete chiral $Z_{2k}$ symmetry
\cite{Kaplan:1984sk}.  Thus by using Wilson fermions in the adjoint representation and tuning the
fermion mass, one  may obtain $N=1$ SYM in the continuum limit
\cite{Montvay:1996ea,Farchioni:2001wx}.  Alternatively, one can
implement chiral fermions on the lattice; this approach has been
explored in
 \cite{Narayanan:1995gw,Huet:1996pw,Nishimura:1997vg,Neuberger:1998bg,Nishimura:1998hu,Kaplan:1999jn,Fleming:2000fa,Fleming:2000hf} 
using domain wall \cite{Kaplan:1992bt,Kaplan:1993sg} or overlap fermions
\cite{Narayanan:1995gw,Neuberger:1998fp} to implement the $Z_{2k}$
chiral symmetry.

In supersymmetric theories with scalar fields, however, there tend to be a plethora
of relevant operators which violate supersymmetry.  There
are no linearly realized  symmetries that can forbid some of
these operators, such as scalar mass terms, other than supersymmetry
itself. Unfortunately, there is no discrete version of supersymmetry
analogous to the lattice subgroup of Poincar\'e symmetry which can be
implemented to 
forbid scalar masses and other unwanted relevant operators, since  supersymmetry
generators are fermionic and there are no macroscopic supersymmetry
transformations.  This suggests that a realization of
supersymmetry in the continuum would greatly benefit by having some subset of the target
theory's  supersymmetry algebra  realized exactly on
the lattice. This has been the recent approach of refs. \cite{Catterall:2001fr,Catterall:2001wx}.

In this paper we discuss a method for implementing supersymmetry on
a spatial lattice with continuous Minkowski time. In particular, we
show that SYM theories with extended supersymmetry in various
dimensions may be constructed on a spatial lattice in a manner that either
 eliminates or  significantly reduces 
  fine tuning problems associated with obtaining the desired target
  theory. 
 Our method is to start with a 
``mother theory'', being a
quantum mechanical system with extended supersymmetry, a large $N$
gauge symmetry, and a global $R$-symmetry group $G_R$ (defined to be
the global
symmetry group of the theory which does not commute with the supercharges).  The spatial
lattice is constructed by ``orbifolding'' by $Z_N$
 gauge and $R$ symmetries, partially breaking the
extended supersymmetry and producing  $N$-site lattice
dimensions. Subsequently taking the continuum limit of this 
``daughter theory''
near a particular point in the classical moduli space of vacua can
result in 
 a higher dimensional quantum field theory with the original extended
 supersymmetry restored, along with 
Poincar\'e invariance.  The advantage of this technique is that the
resultant lattice can retain some exact supersymmetries, which
facilitates recovery of the remaining supersymmetries in the continuum
limit, protecting the theory from undesirable radiative corrections.  


While the present work is confined to examples of spatial lattices
with Minkowski time,
there are no obstructions to extending these techniques to Euclidean
spacetime lattices, the subject of
ref. \cite{Cohen:2003xe} and subsequent papers in preparation.

The  paradigm and motivation for this approach
toward lattice supersymmetry is found in  the deconstruction 
 of SYM theory in 4+1 dimensions, by Arkani-Hamed, Cohen and Georgi
 \cite{Arkani-Hamed:2001ca,Csaki:2001em} 
and the deconstruction of certain 5+1 dimensional theories in
ref.~\cite{Arkani-Hamed:2001ie}.  Our discussion of 
lattices derived from the
mother theory with sixteen supercharges has some overlap with
ref. \cite{Garcia-Compean:1998kh}, and some related results were
derived from string theory by   Arkani-Hamed, Cohen, Karch and Motl (unpublished).

In the next section we  discuss how one can
 start with a ``mother'' theory with a large symmetry group,
and then mod out discrete symmetries (``orbifolding'') to create
daughter theories with a lattice structure.  In the subsequent
section we consider SYM theories with four, eight, and
sixteen supercharges in 0+1 dimensions, and the types of lattices that
can be obtained from them via orbifolding.  We then examine in some
detail the simplest of these lattices, and discuss how to obtain from
it a continuum supersymmetric quantum field theory---in this case,
(2,2) SYM in 1+1 dimensions.   The more complicated, higher
dimensional theories are subsequently treated, although in less
detail.  A summary of supersymmetric notation
relevant for our analysis is 
provided a summary in an Appendix, in order to make this paper more readable.

\section{Orbifolding }
    \label{sec:2}

We begin by describing how one can create a supersymmetric lattice
action by performing an orbifold projection of a supersymmetric
quantum mechanics theory.  This seemingly unnecessarily complicated
approach to constructing a lattice action is justified by the
fascinating and useful properties of the supersymmetric lattices that
result from this procedure.

We start with a mother theory in  
$0+1$ dimensions possessing extended supersymmetry, a  gauge group  $U(k
N^d)$, and a global $R$ symmetry
$G_R$.  By definition, an
$R$-symmetry $G_R$ is a global symmetry under
which the supercharges transform nontrivially.  We will identify the
maximal Abelian subgroup of $G_R$ to be $H_R=U(1)^p$, where $p$ is the
rank of $G_R$. The variables of the daughter theory are then constructed by 
first identifying  a particular $(Z_N)^d$ subgroup of $H_R\times U(k
N^d)$, and then projecting out all fields in the mother theory which
transform nontrivially under this $(Z_N)^d$ symmetry. 
The action of
the theory is simply determined by replacing all of the fields in the
action of the mother theory by their projections. The
resultant action can be thought of as a lattice action, where the
lattice is $d$-dimensional with $N^d$ sites, possessing an independent
$U(k)$ gauge symmetry at each site.   This
technique has been extensively discussed in the
literature in other contexts (see, for example
\cite{Douglas:1996sw,Kachru:1998ys,Schmaltz:1998bg,Strassler:2001fs,Rothstein:2001tu}). The continuum limit of such a
lattice will describe a $d+1$ dimensional quantum $U(k)$ gauge
theory. A special feature of the daughter theory lattice is that  in
the classical continuum limit, all of the supersymmetries of the
mother theory are recovered.  Furthermore, $2^{-d}$ of the mother
theory's supersymmetries are preserved by the orbifold projection, so
that for many of the lattices we construct, there are no relevant
operators allowed which would allow quantum effects to spoil the
classical continuum limit.

To be explicit, consider an 
adjoint field  $\Phi$ in the mother theory, which is a $kN^d\times
kN^d$ matrix, transforming as
$\Phi\to U\Phi U^\dagger$ under the $U(kN^d)$ gauge symmetry. The
field $\Phi$ also transforms as some representation $\CR$ under the
global symmetry $G_R$; it will carry $H_R=U(1)^p$ charges
$(q_1,\ldots,q_p)$ whose values will depend on the representation
$\CR$. From these $q_i$ charges we form $d< p$ linear combinations 
 $\{r_1,\ldots,r_d\}\equiv {\bf r}$, where all $r_a$ charges in the
 theory take on only integer values (we will discuss below how we
 choose these particular combinations of the $q$ charges).  Again, the
 $\bfr$ charges for a given field $\Phi$ 
which will depend upon its particular $G_R$ representation $\CR$.   A
$Z_N^d$ symmetry is then identified, whose action on $\Phi$ is
\beq
\Phi \to  e^{-2\pi i r_a/N}\gamma_a^{-1} \Phi \gamma_a\ ,\quad a=1,\ldots,d.
\label{eq:orbia}
\eeq
Here the  matrix $\gamma_a\in U(k N^d)$ is given by
\beq
\gamma_a = \underbrace{1_N \otimes\ldots}_{a-1}\otimes \Omega\otimes
\underbrace{1_N\otimes\ldots}_{d-a}\otimes 1_k\ , \qquad
\Omega=\rm{diag}(\omega^1,\omega^2,\ldots,\omega^N) \ ,
\eqn{gamdef}\eeq
where $1_N$ and $1_k$ are the $N\times N$ and $k\times k$ unit
matrices respectively, with $\omega = 
e^{2\pi i/N}$.  The orbifold projection then discards all field
components which transform nontrivially under this $Z_N^d$, keeping
only those which satisfy  the $d$ matrix-valued constraints
\beq
\Phi = e^{-2\pi i r_a/N}\gamma_a^{-1} \Phi \gamma_a\ ,\quad a=1,\ldots,d.
\label{eq:orbi}
\eeq

It is convenient to consider $\Phi$ as being composed
of $k\times k$ blocks $\phi_{{\bf mn}}$, where ${\bf m}$ and ${\bf n}$
are $d$-component vectors whose integer components each run from $1$ to $N$.
The orbifold projection results in forcing all of these $\phi_{{\bf
    mn}}$ blocks to vanish except those for which 
\beq
{\bf n} = {\bf m} +
{\bf r}\ .
\eqn{nmr}
\eeq
The surviving $\phi_{\bf  mn}$ blocks satisfying the  condition \Eq{orbi}
are unconstrained, and will be  interpreted as lattice
variables. Thus in the $a^{th}$ subspace, a field $\Phi$ for which
$r_a=0$ will have only diagonal blocks survive, while if $r_a=\pm 1$, 
only the super- or sub-diagonal blocks survive.

\EPSFIGURE[t] {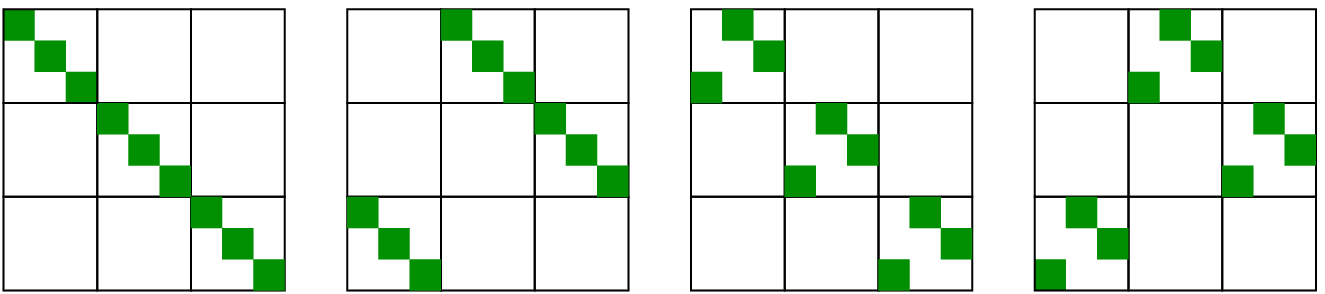,width=11cm}{\sl An example of the
  orbifold procedure in \Eq{orbi} for a $U(9k)$ gauge group with global
  symmetry $G_R=U(1)\times U(1)$,  modded out by $Z_3\times
  Z_3$.  The dark squares are the $k\times k$ blocks  which survive
  the orbifold projection within
  a $U(9k)$ adjoint field with  $G_R$ charges $\bfr $.  The
  examples shown  (from
  left to right in the figure) correspond to $
  \bfr =(0,0)$, $(1,0)$, $(0,1)$ and $(1,1)$. The projected fields may  be interpreted
  respectively 
  as fields living on sites, horizontal links, vertical links, and diagonal links on a
  $3\times 3$ periodic lattice.\label{fig:orbifold}}

When the projected fields satisfying \Eq{orbi} are substituted into
the action of the mother theory, then it becomes a theory of the $N^d$
surviving blocks
$\phi_{{\bf  mn}}$  from each field $\Phi$ of the
mother theory.  This action  is called the daughter theory. The
daughter theory (often called a ``moose'' or a ``quiver'' in the literature)
has a lattice interpretation. The lattice is $d$-dimensional with $N$ sites in each direction, with
each surviving $k\times k$ matrix variable $\phi_{\bf m n}$ residing
on the link between sites ${\bf m}$ and ${\bf n}$   (or at the site ${\bf
  m}$ if ${\bf m}={\bf n}$).  As the mother theory will possess of a
number of different $\Phi$ fields, each with a different ${\bf r}$
charge, the daughter theory will consist of different site and link
variables.  For $d=2$, for example, fields for which ${\bf r}=\{0,0\}$
will live on lattice sites, while those for which ${\bf r}=\{1,0\}$,
$\{0,1\}$ or $\{1,1\}$ will live on horizontal, vertical, or diagonal
links respectively \footnote{If we had started with a mother theory with 
  a $SU(kN^d)$ symmetry instead of a $U(kN^d)$ symmetry, the adjoint
  fields $\Phi$ of the mother theory would have to be traceless,
  resulting in the constraint $\sum_{\bf m}\Tr \phi_{\bf m m}=0$; this
  is equivalent to  a nonlocal constraint on the site variables of the
  resulting lattice, which is awkward to deal with.  It is for this
  reason that we assume a $U(kN^d)$ gauge symmetry in the mother
  theory instead of $SU(kN^d)$.}.  See Fig.~\ref{fig:orbifold} for an example.

The orbifold projection breaks the symmetries of the mother theory.
For example, general $G_R$ transformations which rotate $\Phi$ fields of the
mother theory with different ${\bf r}$ charges into each other are
broken. The $U(kN^d)$ gauge symmetry is also broken by
the orbifold condition \Eq{orbi}, down to a $U(k)^{N^d}$ symmetry.  If
one considers a general $U(kN^d)$ matrix as being composed of $N^{2d}$
$k\times k$ blocks $u_{\bf m n}$, then only those transformations
consisting solely of diagonal blocks ($u_{\bf mn}= u_{\bf m}
\delta_{\bf mn}$) commute with the orbifold condition \Eq{orbi}, where
each  diagonal block $u_{\bf m}$ is an unconstrained time dependent
$U(k)$ matrix associated with the site ${\bf m}$.
Under this unbroken $U(k)^{N^d}$ symmetry, the lattice variables
 transform as $\phi_{\bf m n}\to u_{\bf m} \phi_{\bf m n} u^\dagger_{\bf
  n}$, so that a site variable $\phi_{\bf m m}$ transforms as a
$U(k)_{\bf m}$ adjoint, while a link variable $\phi_{\bf m n}$
transforms as a bifundamental  $(\Yfund,\Ybarfund)$ under $U(k)_{\bf
  m}\times U(k)_{\bf n}$.  This exact $U(k)^{N^d}$ gauge symmetry of
lattice (where the gauge transformations depend only on time) becomes
a $U(k)$ gauge symmetry in the $d+1$ dimensional field theory that
results in the continuum limit, and for each $U(kN^d)$ adjoint field $\Phi$ of the
mother theory, there will result a $U(k)$ adjoint field in the
continuum.  This method of generating a lattice has created the
$d$ spatial dimensions out of the enormous gauge symmetry of the mother theory.

The daughter theory possesses a number of discrete symmetries as
well.  In general, certain discrete $G_R$ transformations
which exchange the $H_R$ charges $q_i$ of the $\Phi$ fields in the
mother theory combine with discrete $U(kN^d)$  symmetries to form the
point group of the resulting lattice (see \cite{Cohen:2003xe} for an
explicit example). Another important discrete symmetry of the lattice
is a $Z_N^d$ symmetry which can be interpreted as the group of
finite lattice translations.  The generators are given by a set of matrices
analogous to the $\gamma_a$ in \Eq{gamdef}, but with the ``clock''
matrix $\Omega$ replaced by an $N\times N$ ``shift'' matrix
\begin{equation}
 T = 
  \begin{pmatrix}
    0\ \  &\ \ 1\ \ &   &       \\
    & \,\,\digc &\,\, \digc &   \\
    &  &    \ \ 0\ \ &\ \ 1\  \\
    1\ \ & &  &\ \ 0\ \\
  \end{pmatrix}\ .
\end{equation}
It is straightforward to see that these translation symmetries commute
with the orbifold condition \Eq{orbi}.  As a consequence of this symmetry,
one sees that our orbifold projection results in a lattice for which
all fields, both fermionic and bosonic, satisfy periodic boundary conditions.  

Finally, the orbifold projection will also break some of the
supersymmetries of the mother theory.  Any gauge invariant operator of
the mother theory carrying nontrivial ${\bf r}$ charge  will
necessarily vanish when the fields out of which it is constructed
satisfy the orbifold condition \Eq{orbi}.   Since by the
definition of an $R$-symmetry, the supercharges of the mother theory
transform nontrivially under $G_R$ and some of these  will  possess
nontrivial ${\bf r}$ charges; such supercharges will therefore will not exist in the daughter
theory. As we will see in the following section, it is possible to
choose the ${\bf r}$ charges such that each of the $d$
$Z_N$ orbifold projections breaks no more than half of the
supercharges of the mother theory; however, it is not possible to
preserve more of the supersymmetry.  As a result, the higher the
dimension of the lattice derived from a given supersymmetric mother
theory, the fewer the number of exact  supercharges it will possess.

All of the cases we will analyze begin with a mother theory which is
equivalent to a super Yang-Mills theory dimensionally reduced from
higher spacetime dimensions down to $0+1$ dimensions.  As such,
neither the mother nor the daughter possess any dimensionful
parameters, other than a factor of $1/g^2$, where $g$ is the gauge
coupling constant, which multiplies the entire action.  Thus even
though the daughter theory has a lattice structure, the lattice cannot
be identified as a spacetime lattice at this point, since there is no
parameter that can be identified as a lattice spacing $a$, nor any
``hopping'' terms to allow for spatial propagation.
  The hopping terms are
obtained in deconstruction models by giving vevs to the bifundamental
scalar fields living on the links, where the vev plays the role of the
inverse lattice spacing
\cite{Arkani-Hamed:2001ca,Arkani-Hamed:2001ie}. In supersymmetric
field theories, the vevs are chosen to lie along the flat directions
(moduli) of the classical ground vacuum. In many cases of interest, these flat
directions are known to persist in the quantum theory. However, in
supersymmetric quantum mechanics, the flat directions in the classical
groundstate energy never persist when quantum corrections are
included.  As we will show, however, at least in weak coupling,
the classical flat directions are expected to become quantum
mechanically flat in the continuum limit.  As we will discuss in
\S\ref{sec:4b}, it is possible to localize the moduli sufficiently
well that they can be treated as classical variables provided that the
spatial lattice is sufficiently large, and we do not look at
correlation functions over too long a time.

The lattices obtained by this  orbifolding and moduli fixing procedure possess a number of features
important obtaining a supersymmetric and Lorentz invariant continuum
limit:
\begin{enumerate}[{(i)}]
\item
  The daughter theory is a $d$-dimensional lattice with a $Z_N^d$
    symmetry, which will become the group of spatial translations  in
    the continuum limit.  The point group of the lattice is a
    precursor to both the Lorentz symmetry and the non-Abelian
    $R$-symmetry of the continuum theory;
\item
  The lattice can be exactly supersymmetric, which greatly reduces the
  number of relevant operators to tune in order to recover the full
  supersymmetry and Poincar\'e group of the target theory;
\item
  There are no unwanted fermion ``doubler'' modes which need to be
  removed by additional Wilson operators, which would spoil the
  supersymmetry --- all of the
  fermion modes that survive the continuum limit play a needed role in
  the target theory;
\item
  Gauge fields appear as bifundamental link variables, unlike in the
  Wilson implementation of lattice gauge symmetry; this allows a
  supersymmetric treatment of gauge bosons and their gaugino partners,
  which also transform as link variables.
\end{enumerate}

\section{Three mothers and six daughters}
\label{sec:3}

 In the examples we shall discuss, the mother theory is $0+1$
 dimensional SYM with four, eight, or sixteen supercharges.  These
 theories may be
derived by dimensional reduction to 0+1 dimensions of $N=1$ SYM theory from 3+1, 5+1 or
9+1 dimensions respectively.  In those dimensions, each of the  $N=1$ SYM
theories consists of only gauge bosons and gauginos;  however, when
reduced to lower dimensions, the extra gauge boson polarizations
become scalars, and the extra gaugino components become extra
fermions, so that the resulting theory is a gauge theory with matter
fields in the adjoint representation with Yukawa and $\phi^4$ interactions, whose strengths
are all related to the gauge coupling. The $R$-symmetry group  $G_R$ of
these theories is just a product of the original
$R$-symmetry before dimensional reduction and the rotation
group of the eliminated spatial dimensions.

In the first example,  N=1 SYM in
3+1 dimensions has a  $U(1)$  $R$-symmetry, so that when dimensionally
reduced to $0+1$ dimensions, the theory has $G_R= 
SO(3)\times U(1)$.  The field content of the dimensionally reduced
mother theory consists of four bosons (the gauge fields of the $3+1$
dimensional theory) transforming under $G_R$ as $1_0\oplus 3_0$,
and four real fermions (the original gaugino fields) transforming as
$2_1\oplus 2_{-1}$.

The second example, N=1 SYM in
5+1 dimensions, has an $SU(2)$ 
$R$-symmetry, so that after dimensional reduction to $0+1$ dimensions,
the $R$-symmetry is $G_R=
SO(5)\times SU(2)$.  In this case the original theory had a six
component gauge field, and a complex, four component Weyl fermion (the gaugino)
transforming as an adjoint under the gauge symmetry; in the dimensional reduced
mother theory, the $v_0$ gauge field transforms as a singlet under
$G_R$, while the vector potential $\vec v$ transforms as the $(5,1)$ representation, and
the eight real fermionic degrees of freedom transform as a $(4,2)$.\

Finally,  N=1 SYM in
9+1 dimensions possesses no 
$R$-symmetry, so that when dimensionally reduced to $0+1$ dimensions $G_R=
SO(9)$.  The bosons transform as a $9\oplus 1$, while the
sixteen real fermions are in the $16$ dimensional spinor
representation of $G_R$.

%
%
\TABULAR[t]{ccccc}{
$N_{\rm SUSY}$ & $G_R$ & Rank & fermions &  bosons 
\\
\hline  &&&&\\
4& $ U(1)\times SO(3)$ & 2  & $(\pm\half,\pm\half)$ & $2(0,0)\oplus (0,\pm1)$ \\ &&&&\\
8& $ SU(2)\times SO(5)$ & 3  & $(\pm\half,\pm\half,\pm\half)$ & 
$2(0,0,0)\oplus (0,\pm1,0)\oplus (0,0,\pm1)$ \\ &&&&\\
16& $SO(9)$ & 4  & $(\pm\half,\pm\half,\pm\half,\pm\half)$ &
$2(0,0,0,0)\oplus (\pm1,0,0,0)$\\ &&&& $\oplus (0,\pm1,0,0) \oplus
(0,0,\pm1,0)$\\ &&&&$ \oplus (0,0,0,\pm1)$ \\ \hline&&&&\\ 
 }{ Properties of $N=1$ SYM theories  
  from $d=3+1$, $d=5+1$, and $d=9+1$ dimensions, dimensionally reduced
  to  
  $d=0+1$. Listed are $N_{\rm SUSY}$ (the number of  
  real supercharges); the resultant $R$-symmetry $G_R$ and its rank; and
  the  charges $(q_1,q_2,\ldots)$ of the  fermions  and bosons
  under the maximal
 Abelian subgroup $H_R\subset G_R$. In each of these theories, the
 supercharges carry the same $H_R$ charges as do the fermions.\label{tab:table1}}
 In Table~\ref{tab:table1} we display the $R$ symmetry
group $G_R$ in each of the three cases we examine, dimensionally
reduced SYM theories from $3+1$, $5+1$ and $9+1$ dimensions with four,
eight and sixteen real supercharges respectively.  We have
chosen a convenient basis for $H_R$, the maximal Abelian subgroup of
$G_R$, classifying the $U(1)$ $H_R$ charges
of the bosons
and fermions as $(q_1, q_2,\ldots)$, where the number of $q_i$
charges equals the rank of $G_R$.  The orbifold charges $\bfr $ in
\Eq{orbi} will then be taken to be linear combinations of these $q_i$.

How exactly are we to define the $\bfr$ charges?  There are several
requirements on the definition of $\bfr$:
\begin{enumerate}[{(i)}]
\item  Each component $r_a$ must be a linear combination of the $q_i$
  charges;
\item Each component of $r_a$ must assume only integer values, since
  we require that $e^{2\pi i r_a/N}\in Z_N$;
\item The simplest lattice with only nearby interactions requires that
  the $r_a$ only take on values equal to $0$ or $\pm1$ (see \Eq{nmr});
\item One unbroken supersymmetry survives the orbifold projection for
  each fermion with $\bfr=0$, and so the $\bfr$ charges should be
  defined so as to maximize the number of $\bfr=0$ fermions.
\end{enumerate}
The last point follows because the supercharges for the theories in
Table~\ref{tab:table1} share the same $G_R$ representation as
 the fermions.  As mentioned in the previous symmetry, a gauge
invariant operator with $\bfr\ne0$ does not survive the orbifold
projection.  Therefore we wish to maximize the number of fermions with
$\bfr=0$ in order to obtain the most supersymmetric lattice possible.
It is easy to see that with the fermion charges shown in
Table~\ref{tab:table1}, at most half of the fermions can be neutral under
any linear combination of the $q_i$ charges.  Thus each $Z_N$ orbifold
projection must break at least half of the remaining supercharges. 

It is easy to understand the difference between the  supersymmetries which
survive the projection and those which do not. Since the unbroken
supercharges have $\bfr=0$, they interchange bosons and fermions at
each site or at each link on the lattice.  The broken supercharges are those
with nonzero $\bfr$ charge;  they exchange bosons and fermions in the
mother theory which end up at different locations on the lattice.  For
example, they might exchange a boson at a  site with a fermion
at a link.  While these latter transformations are not exact
symmetries of the lattice, they become exact in the continuum limit
when the distinction between site and link variables becomes
irrelevant. 

All four of the above requirements on the $\bfr$ charge assignments are most simply met if we
define 
\beq
r_a = (q_1-q_{a+1})\ ,\quad a=1,\ldots,d\ .
\eeq 
 We now examine the various
lattices that can result from  the
theories in Table~1 with this definition of $\bfr$.

\subsection{One dimensional lattices}
\label{sec:3a}

The simplest one dimensional lattice we can construct follows from
$r\equiv(q_1-q_2)$, which takes on the values $0$ or $\pm 1$ for every
field in Table~1.
  Fig.~\ref{fig:lat1d} displays the lattice obtained by  orbifolding
  by the single $Z_N$ factor.  This one-dimensional spatial lattice with
  $N$ sites
will preserve half the supersymmetries of the mother theory.  The
matter content on the sites and links can be read off of
Table~\ref{tab:table1}, given that fields 
with $r=q_1-q_2=\pm 1$ live on the links, while the remaining fields
have $r=0$ and live on the sites.  For example, the simplest of our
mother theories contains four bosons, four fermions, and four
supercharges (all real); after orbifolding by $Z_N$ we get the lattice of
Fig.~\ref{fig:lat1d} with two exact supercharges, two bosons plus two
fermions at each site, and  two bosons plus two
fermions on each link. This particle content
corresponds to a vector supermultiplet at each site, and a bosonic chiral
supermultiplet on each link. (See Appendix~A for a summary of the relevant supermultiplet
structure for $0+1$ dimensional supersymmetry.)
 The target theory in the continuum limit in
this case will be the $1+1$ dimensional SYM theory with $(2,2)$
supersymmetry.  We will consider the continuum limit of this theory
more closely in \S\ref{sec:4}.

In the case of the mother theory with eight supercharges, there will
be four fermions and four bosons on each site, with two real bosons and
four real fermions on each link, and four exactly conserved
supercharges.  The
mother theory with sixteen 
supercharges gives rise to the one-dimensional lattice with six real
bosons and eight real fermions at each site, 
four real bosons and eight real fermions on each link, and eight exact
supercharges; it is thought to be a trivial theory in the infrared \cite{Seiberg:1998ax}.

\DOUBLEFIGURE[t]{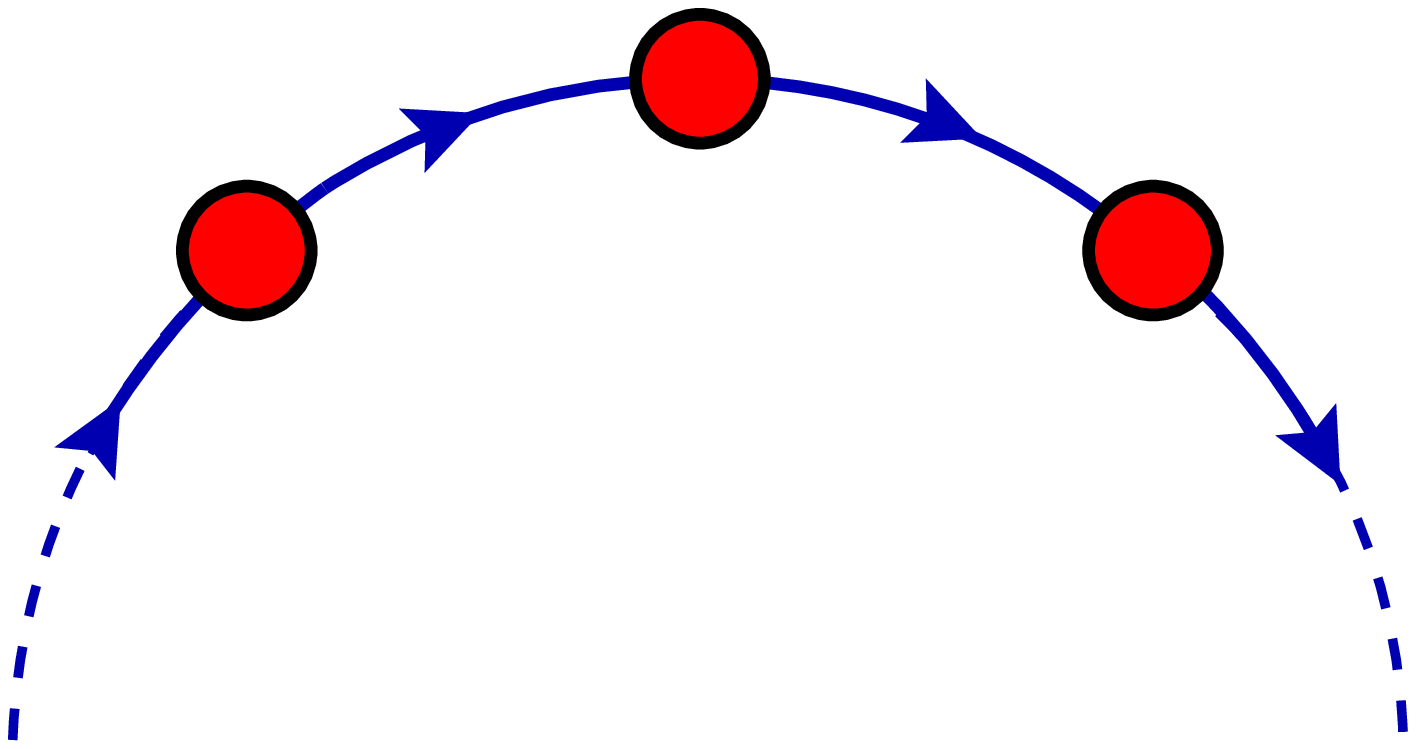,width=6cm}{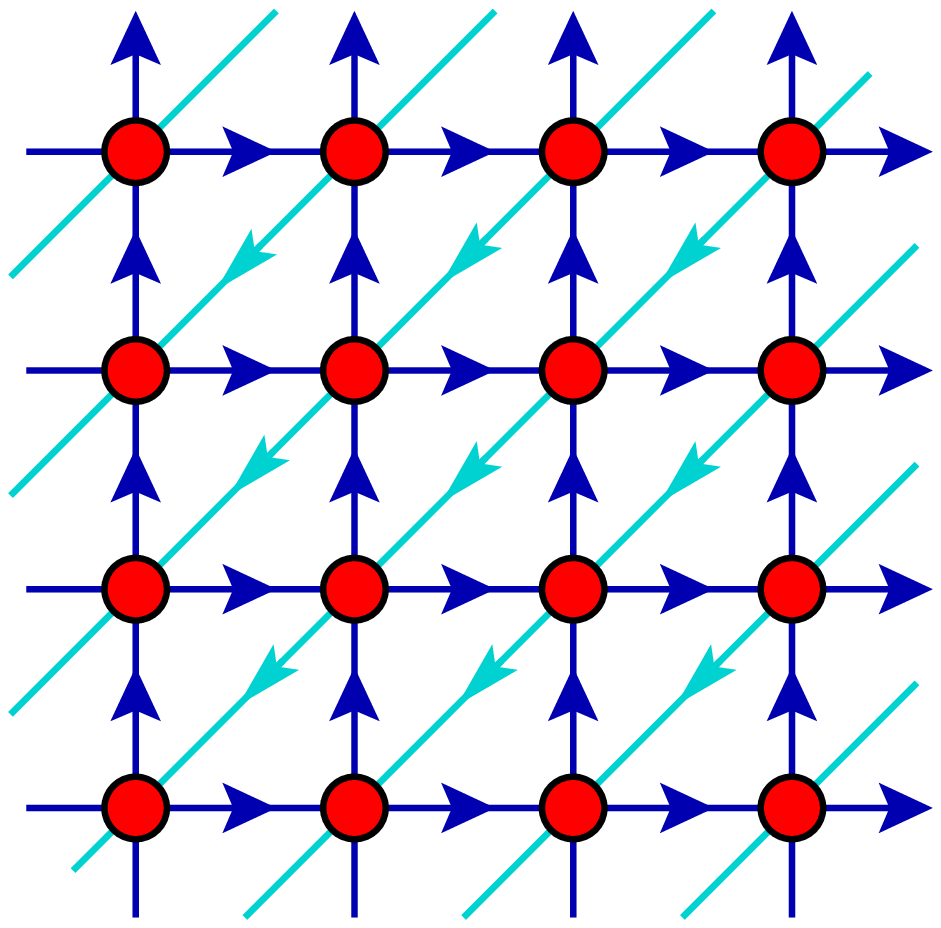,width=6cm}
{\sl The lattice  with $N$ sites obtained by 
   orbifolding a SUSY QM  theory with a $U(kN)$ gauge symmetry by
   $Z_N$. 
   Sites correspond to $U(k)$ gauge groups with adjoint matter,  while links
   correspond to matter superfields  transforming as
   bifundamentals.\label{fig:lat1d}}
{\sl The two dimensional lattice preserving two exact supercharges
   obtained by 
   orbifolding the eight supercharge $0+1$ SYM theory by  
   $(Z_N\times Z_N)$.  Sites correspond to  vector supermultiplets
   transforming as $U(k)$ adjoints,  dark links represent
   bifundamental bosonic chiral supermultiplets, while
   light links are bifundamental Fermi supermultiplets
   . \label{fig:lat2dsquare} }

\subsection{Two dimensional lattices}
\label{sec:3b}

Mother theories with eight or sixteen supersymmetries allow
for the orbifolding by a $Z_N\times Z_N$ symmetry.  For these lattices
we choose
\beq
r_1=  (q_1-q_2)\ ,\qquad
 r_2= (q_1-q_3)\ .
\eqn{r28}
\eeq

Starting from the mother theory with eight supercharges, the
$Z_N\times Z_N$ orbifold with the above choice for the ${\bf r}$
charges  reduces  the exact
supersymmetries from eight to two, and results in the two
dimensional lattice shown in  Fig.~\ref{fig:lat2dsquare}.
 At each site one finds two real bosons and two
real fermions corresponding to the vector supermultiplet in 0+1
dimensions with two supercharges;  the horizontal and vertical links
each represent two real bosons and two real fermions, which constitute
a bosonic chiral multiplet; on the diagonal links are the final two
real fermions and no bosons, which comprise a Fermi
multiplet. The action (discussed
in \S\ref{sec:5a}) involves
both kinetic terms and superpotential 
terms which are triangular plaquette interactions. The point symmetry group
of the action consists reflections about the diagonal axes and rotations by $\pi$; together these
transformations generate the four element group $C_{2v}$. The target
theory in this case, a $2+1$ dimensional SYM theory with eight
supercharges, has been discussed in \cite{Seiberg:1996bs,Seiberg:1999xz}.

\DOUBLEFIGURE[t]{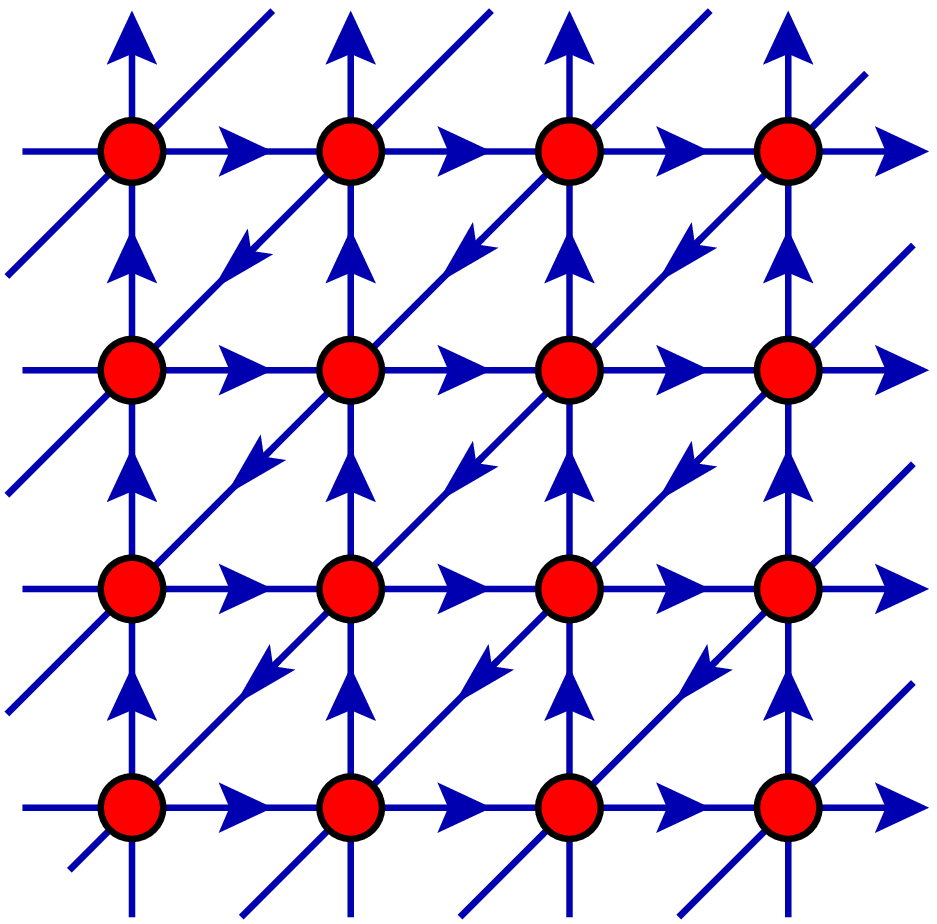,width=6cm}{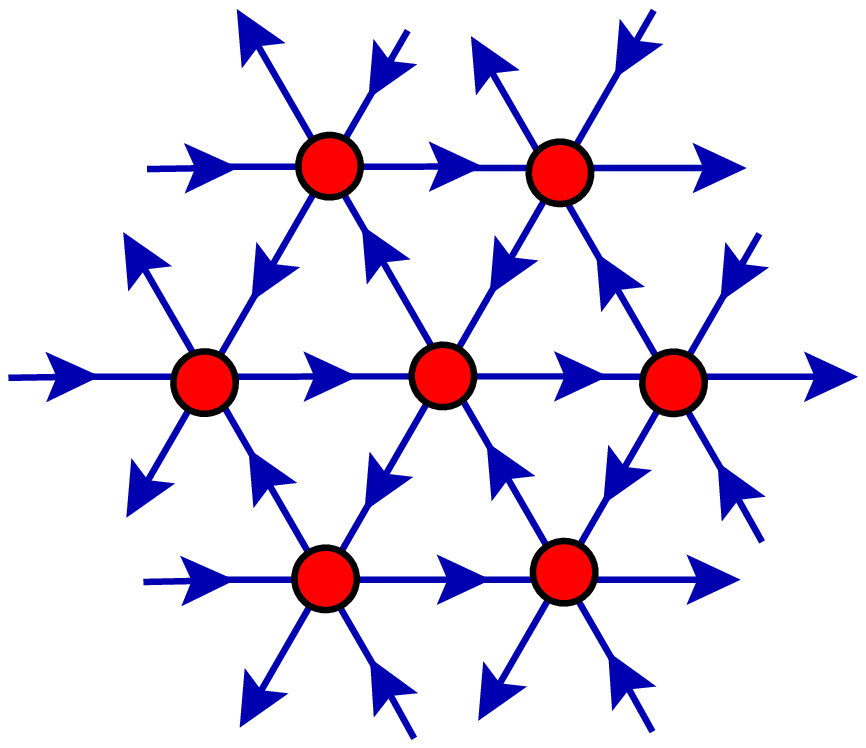,width=7cm}
{\sl  When the sixteen supercharge $0+1$ SYM theory is orbifolded by  
   $(Z_N)^2$, one obtains a similar lattice to that in Fig.~3, except
   that four supercharges are preserved, and the $\hat x$, $\hat y$
   and diagonal links all correspond to
   identical  supermultiplets, with one complex scalar and two complex
 one-component fermions. \label{fig:lat2dsquare16q} }
{\sl The lattice of Fig.~4 hides the fact that the two dimensional
  lattice with two exact supercharges is invariant under a $D_{6h}$
   discrete symmetry preserved by the orbifold projection. It is
   therefore appropriate to  represent the theory instead by 
   the above hexagonal lattice. \label{fig:lat2dhex} }%

{\ }

\DOUBLEFIGURE[t]{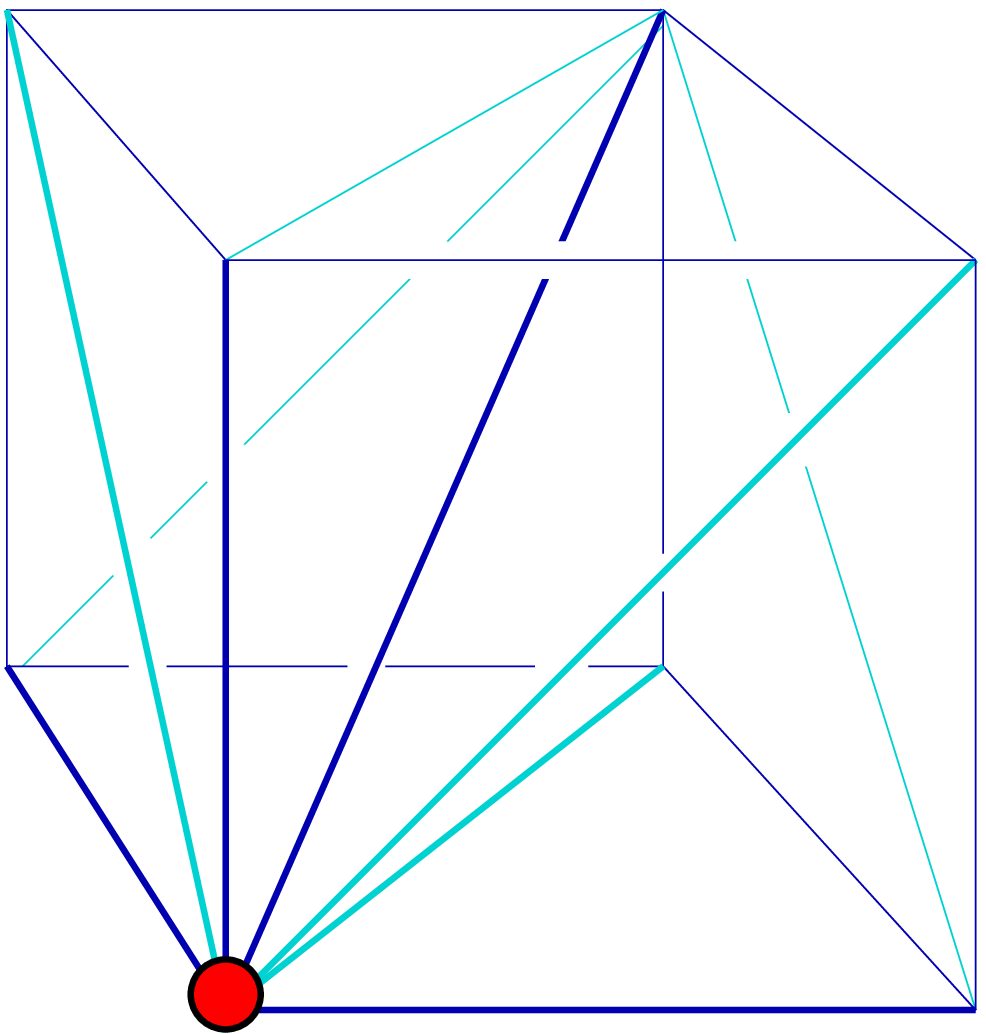,width=5cm}{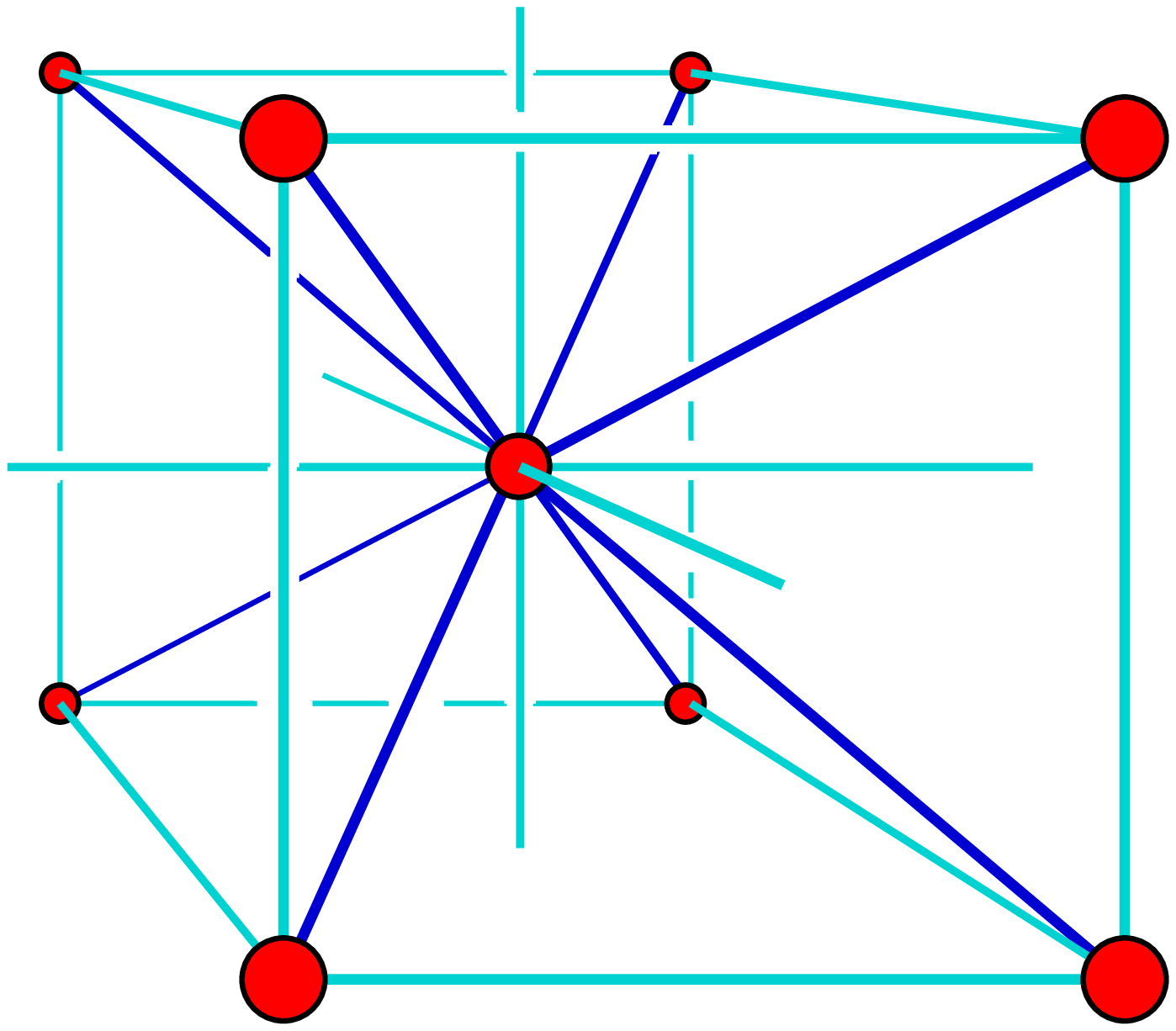,width=7cm}
{\sl The sixteen supercharge $0+1$ SYM theory  orbifolded by  
   $(Z_N)^3$ results in a  lattice  respecting
   two exact supercharges with the above structure.  The target theory
   is $N=4$ SYM theory in $3+1$ dimensions. Dark
   blue links correspond to bosonic chiral superfields, light blue
   links are  Fermi superfields, and a vector supermultiplet
   resides at each site. \label{fig:lat3dcube} }%
{\sl The lattice of Fig.~6 hides the $O_h$ symmetry of the action, 
  which  is better represented by the above
    body-centered cubic crystal.  Dark lines are
  bosonic chiral superfields, while light lines are 
  Fermi superfields; triangle plaquette terms appear in the
  superpotential (link orientation not displayed). \label{fig:lat3dBCC}}
{\ }
When the mother theory has sixteen supercharges, the lattice we obtain
from the ${\bf r}$ charges \Eq{r28} is shown in
Fig.~\ref{fig:lat2dsquare16q}. The lattice retains four exact supercharges,
there are four bosons and four real fermions at each site (a vector supermultiplet), and two
real bosons and four real fermions on each link (a chiral
supermultiplet). The chiral multiplets get  charges $(r_1, r_2)$ as $(1,0), 
(0,1), (-1,-1)$. Thus the diagonal multiplet is also a bosonic chiral 
multiplet (as opposed to a Grassmann multiplet.) There is a symmetry permuting 
these chiral link variables, and so the symmetry group of the daughter
theory is larger than it looks like in Fig.~\ref{fig:lat2dsquare16q}. Hence 
we draw it as an hexagonal lattice as in Fig.~\ref{fig:lat2dhex},
which makes manifest the point group symmetry of the 
action the 12 element 
$D_6$ dihedral symmetry group. The target theory in this case, a $2+1$ dimensional 
field theory with sixteen supercharges, is expected to be quite
interesting with an interacting superconformal phase and an enhanced
$R$-symmetry related to the $SL(2,Z)$ symmetry of $N=4$ SYM in $3+1$
dimensions \cite{Seiberg:1998ax}. We discuss this lattice further in
\S\ref{sec:5b}.

 \subsection{The three dimensional lattice}
\label{sec:3c}

As our last example, one may orbifold the theory with sixteen
supercharges by $Z_N^3$ and still retain two exact
supersymmetries. The target theory in this case is $N=4$ SYM in $3+1$ dimensions.
Following the prescription described above, we choose the $\bfr $
charges for the orbifold to be
\beq
r_1=(q_1-q_2)\ ,\quad
r_2=(q_1-q_3)\ ,\quad
r_3=(q_1-q_4)\ ,
\eqn{r316}\eeq
This orbifold condition projects the four bosonic chiral multiplets along the
dark blue  links of Fig~\ref{fig:lat3dcube}, and three Fermi multiplets 
along the light  blue  links of Fig~\ref{fig:lat3dcube}.  A vector multiplet resides
at each site.  As in the case of the two dimensional case, the action
is invariant under a larger discrete symmetry  than is apparent from
the lattice defined by the ${\bf r}$ vectors. This is the  octahedral
symmetry with 
inversions, the 48 element group $O_h$. Such a symmetry is made
possible by the fact the the superdiagonal link and the cube edge
links correspond to variables in the same supersymmetry multiplets.
The action is more faithfully described then by the body centered
cubic lattice of Fig.~\ref{fig:lat3dBCC}.  Note that the connectivity
of the two lattices is identical.
The action for this lattice is  discussed in \S\ref{sec:5c}.

\section{The 1+1 dimensional theory with $(2,2)$ supersymmetry}
\label{sec:4}

\indent
A key feature in how orbifolding is able to preserve some lattice
supersymmetry is  the manner with which it treats gauge bosons and the
associated gauginos of the target theory. Rather than having gauge
fields appear as unitary matrices on  links in the manner of Wilson,
with gauginos represented as adjoints at the sites, the lattices
discussed here have both fermion and boson bifundamental
representations living on the links, making a symmetry between them
possible. As a result, however, we are forced to think more carefully
about the continuum limit, since the lattice spacing is no longer a
parameter of the theory, but is rather dynamical, associated with the
classical value of the bosonic fields about which there are quantum
fluctuations.

\subsection{The lattice action and continuum limit for (2,2) supersymmetry in $1+1$
  dimensions}
\label{sec:4a}

In this section we work through our simplest example,
whose target theory is $(2,2)$ supersymmetric Yang-Mills theory in
$1+1$ dimensions. This theory is just the dimensional reduction to
$1+1$ dimensions of
$N=1$ SYM theory in  3+1 dimensions, written in component fields
as\footnote{We follow the ``mostly minus'' metric convention $\eta_{\mu\nu}={\rm
    diag}(1,-1)$, and all $U(k)$ generators are normalized as \hbox{$\Tr T_a
T_b=\delta_{ab}$.}}
\beq
\CL_{1+1} = \frac{1}{g_2^2}\Tr\left(-\frac{1}{4} F_{\mu\nu}F^{\mu\nu}
 +\mybar\Psi 
  i\sla{D}\Psi + (D_\mu S)^\dagger (D^\mu S) + \sqrt{2}(\mybar \Psi_L
  [ S,\Psi_R] + h.c.) - 
 \frac{1}{2}[  S^\dagger, S]^2\right)
\eqn{targ}
\eeq
where $ S$ is a complex
scalar and $\Psi$ is a two component Dirac fermion, each  transforming
as an adjoint under the $U(k)$ gauge symmetry, with gauge bosons $(v_0,v_1)$.  Note 
that while $U(k)$ is not a simple group, and in principle can have
two independent coupling  constants, we are setting them to be equal.
In the continuum, the $U(1)$ fields are all
noninteracting and 
decouple. The apparently problematic existence of a free massless 
scalar in this model (the  partner of the photon) will be resolved
below.     This is a super-renormalizable theory, which would exhibit
logarithmic divergences for the scalar mass, were it not for
supersymmetry which causes both the infinite and finite contributions
to cancel.

The 0+1 dimensional lattice theory of Fig.~\ref{fig:lat1d} with two supercharges can be
conveniently  expressed using the superfield
notation  of dimensionally reduced $(0,2)$ supersymmetry in $1+1$
dimensions, as described in Appendix~A. There are two types of
superfields in this model: vector superfields $V$ residing at each
site,  containing a gauge field $v_0$, a complex fermion $\lambda$,
a real scalar $\sigma$, and a real auxiliary field $d$; and a chiral
superfield $\Phi$ on each link, 
consisting of a complex scalar $\phi$ and a complex fermion
$\psi$. All fields are $k\times k$ matrices, with $\lambda_n$ and
$\sigma_n$ transforming as adjoints under the $U(k)_n$ gauge
symmetries, and $\phi_n$ and $\psi_n$ transforming as bifundamentals
$(\Yfund,\Ybarfund)$ under the $U(k)_n\times  
U(k)_{n+1}$ gauge symmetries associated with the link ends.

In terms of these superfields, the lattice theory is given by
\beq
S = \frac{1}{ g^2}\int\, dt\, \sum_{n=1}^N\Tr\left[\frac{1}{2}\mybar\Phi_n i \CD^-_0 
\Phi_n + \frac{1}{8}\mybar\Upsilon_n \Upsilon_n 
\right]_{\mybar\theta\theta}\ .
\eeq
where $\Upsilon_n$ is the Grassmann chiral multiplet containing the gauge
kinetic terms at site $n$.  We have
periodic boundary conditions with
$\Upsilon_{N+1}\equiv \Upsilon_1$.
After eliminating the auxiliary fields $d_n$, an expansion in terms of
component fields yields
\begin{eqnarray}
L = \frac{1}{g^2}\sum_n \Tr &\Bigl[&\frac{1}{2}(D_0\sigma_n)^2+
  \,\mybar{\lambda}_n \, iD_0 \,\lambda_n
 + |D_0\, \phi_n|^2 +   \mybar{\psi}_n\, i D_0\psi_n \Bigr.\cr
&&
- \mybar{\lambda}_n[\sigma_n , \lambda_n]+
 \mybar{\psi}_n(\sigma_n  \psi_n-\psi_n\sigma_{n+1}) -
 \sqrt{2}\left(i\mybar\phi_n(\lambda_n\psi_n+\psi_n\lambda_{n+1}) +
   h.c.\right)
\cr
&&
\left.
-\left| \sigma_n\phi_n-\phi_n\sigma_{n+1}\right|^2
-\frac{1}{2}\left(\phi_n\mybar\phi_n-\mybar\phi_{n+1}\phi_{n+1}\right)^2\right]\
,  
\eqn{tcomp}
\end{eqnarray}
where
\beq
D_0\phi_n = \partial_0\phi_n + iv_{0,n}\phi_n -i\phi_nv_{0,n+1}
\eeq
and similarly for $D_0\psi_n$.

The theory has  a classical moduli
space corresponding to  all the $\phi_n$
being equal and diagonal, up to a gauge transformation.  We now choose
to expand about the point in the classical moduli space which
preserves a $U(k)$ symmetry, namely
\beq
\phi_n = \frac{f}{\sqrt{2}}\times {\bf 1}_k\ ,
\eeq
where $ {\bf 1}_k$ is the $k\times k$ unit matrix and $f$ is a
constant parameter with
dimensions of mass.  We then define a lattice spacing $a$,
compactification scale $L$ and gauge
coupling $g_2$ by \footnote{Note that with our normalizations,
  quantities have the following mass 
dimension:
$$
[\,{\rm bosons}\,]=[\,f\,]=[\,a^{-1}\,]=1\ ,\quad [\,{\rm fermions}\,]=3/2\ ,\quad
[\,g^2\,]=3\ ,\quad [\,g_2^2\,]=2 \ .
$$}
\beq
a\equiv \frac{1}{f}\ ,\qquad L\equiv N a\ ,\qquad g_2^2\equiv a g^2\ .
\eeq
The target theory is obtained by taking the continuum limit $a\to 0$
and $N\to \infty$ for fixed $L^2 g_2^2$, such
that the dimensionless quantity $ a^2 g_2^2$  goes to zero:
\beq
a^2 g_2^2=a^3 g^2 \to 0\ .
\eqn{cont}
\eeq
It follows that $a/L\to 0$ as well. The discrete sum  $\sum_n$ is
replaced by a spatial integral $\int 
\frac{{\rm d}x}{a}$, finite differences are replaced by
a derivative expansion, such as
\beq
\phi_{n+1}(t)-\phi_n(t)\longrightarrow a\partial_x\phi(x,t)
+\frac{1}{2}a^2\partial_x^2\phi(x,t)+\ldots\ .
\eeq
It is convenient to write $\phi$ in terms of two hermitian fields
$h_{1,2}$ as
\beq
\phi(x,t) = \frac{h^{(1)}(x,t) + i h^{(2)}(x,t)}{\sqrt{2}}\ .
\eeq
Then one finds that at the classical level one obtains the
target theory \Eq{targ} in the continuum limit \Eq{cont}, after making
the
identifications between continuum and lattice variables
\beq
v_0=v_0\ ,\quad v_1= h^{(2)}\ ,\quad
 S = \frac{\sigma +i h^{(1)}}{\sqrt{2}}\ ,\quad
\Psi =
\begin{pmatrix}
\psi\cr \mybar\lambda
\end{pmatrix}\ ,
\eeq
where $\Psi$ is given in the Dirac basis, 
\beq
\gamma^0=
\begin{pmatrix} 1
  & \ \ 0 \cr 0 & -1\cr
\end{pmatrix}\ ,\qquad 
\gamma^1 =
\begin{pmatrix}
0 & -1\cr   1 &\ \ 0
\end{pmatrix}\ ,\qquad
\gamma_5 =\gamma^0\gamma^1=-
\begin{pmatrix}
0 &\  1\cr   1 & \ 0
\end{pmatrix}\ .
\eeq
The gauge field $v_0$
is  the same in both the lattice and the 
target theories. There are no unwanted fermion
doublers in the spectrum;  note that the $\mybar \phi \psi \lambda$
interaction in \Eq{tcomp} with the substitution of $\mybar\phi\to f/\sqrt{2}$ 
 looks like the standard kinetic term for a Wilson fermion with $r=1$.

The infinite volume limit may then be taken by sending $L^2 g_2^2\to
\infty$. 
In doing so, we must take care to treat separately the zeromode
corresponding to shifts in the scale $f=1/a$:
\beq
\zeta(t) \equiv \frac{1}{\sqrt{Nk}}\sum_n\Tr h^{(1)}_{n}(t)\ .
\eeq
We will refer to this mode as the ``radion'', since a shift in $\zeta$
corresponds to  shift in the lattice size $L$ (and the lattice
spacing). With this radion mode treated separately, the infrared divergence in the propagator of
the  gauge singlet 
piece of the $S$ field in the target theory scales as $\ln
L/a=\ln N$ rather than being infinite.  As we will argue, so long as we specify the limits
\Eq{cont} such that $(a g_2)^2\ln N\to 0$, this infrared
divergence should be harmless (see Ref.~\cite{Cohen:2003xe} for a more
detailed discussion). 

The theory that results at finite lattice spacing $a$ is the target
theory \Eq{targ}, with the addition of operators that vanish as powers
of $a$.  Since some of these operators do not respect the symmetries
of the target theory,  care
must be taken to analyze renormalization as $a\to 0$. Another issue to
address is the role of the radion $\zeta(t)$---it is well known that
the degeneracy of a classical moduli space is lifted in quantum
mechanics.  For example, a classical diatomic molecule has a continuous
ground state degeneracy corresponding to its angular orientation; in
quantum mechanics, however, that degeneracy is lifted and the unique
ground state has vanishing angular momentum, $\ell=0$.  One can think
of the quantum fluctuations in the molecule's orientation as
``destroying'' the classical ground state.  Similarly, one might well
worry that fluctuations in $\zeta$ (and hence the
lattice spacing) will destroy
the spatial interpretation of our lattice.  We will return to this
worry in \S\ref{sec:4c} and show it is unfounded, under certain
restrictions on how the lattice is used.

\subsection{Renormalization}
\label{sec:4b}
We ignore the zeromode $\zeta(t)$ for the moment, and consider
renormalization of the target theory as $a\to 0$.  Aside from the
desired operators \Eq{targ}, we also find at finite $a$ operators
which are gauge and translational invariant, but which  respect
neither 1+1 dimensional Lorentz symmetry nor the enhanced $(2,2)$
supersymmetry. For example, when the fields are canonically normalized
one finds the marginal  operators such as
\beq
\Tr (\phi\partial_1\mybar\phi)^2\ ,\qquad \Tr
\mybar\psi\partial_1\psi\phi\ .
\eqn{badops}
\eeq
At first sight operators like these are  a disaster;  however when
fields are canonically normalized they have coefficients $(g_2 a)^2$
and $(g_2 a)$ respectively, and even when inserted in loops they prove
to be harmless due to the exact supersymmetry of the underlying lattice.

The simplest way to analyze this theory is to maintain the
noncanonical normalization used in \Eq{targ}, with a factor of
$1/g_2^2$ in front of the effective action.  At
tree level and finite lattice spacing $a$, the effective action has an overall factor of
$1/g_2^2$, and  therefore by dimensional analysis, operators of 
the form $\phi^a \partial^b \psi^{2c}$ come with  coefficient
proportional to $a^{p-4}$, where $p=(a+b+3c)$
(here $\phi$ is a generic boson field, $\psi$ is a fermion field, and
$\partial$ signifies either an $x$ or $t$ derivative anywhere in the
operator). the generic form of operators with $p\le 4$ are given in Table~2.

%
%
\setlength{\extrarowheight}{5pt}
\TABULAR[t]{|c|c|}{
\hline
$p=a+b+3c$ & $\phi^a\partial^b\psi^{2c} $
\\
\hline \hline 0 & 1 \\
\hline 1 & $\phi$ \\
\hline 2 & $\phi^2$ \\ 
\hline 3 & $\phi^3$, $\psi\psi$, $\phi\partial\phi$  \\ 
\hline 4 & $\phi^4$, $\phi\psi\psi$,  $(\partial\phi)^2$, $\psi\partial\psi$,
$\phi^2\partial\phi$  \\ \hline 
 }{Engineering dimension of operators in $d+1$ dimensions,  for the field
 normalization that gives an overall factor 
 $1/g^2$ in the action. $\phi$ and $\psi$ are generic bosons and
 fermions respectively; $\partial$ corresponds to a derivative.\label{tab:table2}}

 Explicit
calculation shows that at tree level all 
operators with  $p> 4$ have vanishing coefficients in the $a\to 0$ limit, while the $p=4$
operator coefficients  have precisely the values of the
 target theory \Eq{targ}.  
Perturbative quantum effects at $\ell$ loops will generically shift the
coefficient of the operator  $\phi^a \partial^b \psi^{2c}$ by
 $a^{p-4} (a g_2)^{2\ell}$ times possible
logarithms of the form $\ln(L/a)=\ln N$.
We see that  only operators with $p\le 2$ can receive
divergent renormalization at one loop, and only the cosmological
constant ($p=0$) can be infinitely renormalized at two loops.  All higher
loop graphs are finite, as befits a super-renormalizable theory. Thus
for the target theory to be 
obtained in the $a\to 0$ limit requires that the radiative
corrections  which violate either $1+1$ dimensional Lorentz invariance
or $(2,2)$ supersymmetry correspond only to operators with
$p> 2$. 

Our task is then to consider the dangerous operators with
$0\le p\le 2$ and ask whether these operators are consistent with the
exact symmetries of the underlying $0+1$ dimensional lattice.  if they
are not, then they cannot be generated as radiative corrections in our
theory.
From Table~2, we see that the potentially dangerous operators are: (i)
$p=0$: a cosmological constant; (ii) $p=1$: a boson tadpole $\phi$; (iii)
$p=2$: a boson mass $\phi^2$.
 It is straightforward to see that in fact the symmetries of the underlying $0+1$
dimensional supersymmetric lattice forbid both the $\phi$ and the
$\phi^2$ operators, as there are no Fermi multiplets in the spectrum,
and hence no way to introduce a superpotential. However, it is possible to include a
    Fayet-Iliopoulos term 
\beq
\frac{i}{g^2} \frac{\xi}{N}\sum_n\Tr \Upsilon_n + {\rm h.c.}
\eeq
for the $U(1)$ field strength, the effect of which is solely to
contribute to   the $1+1$ dimensional
theory  a cosmological constant proportional to $\xi^2$, corresponding to the
operator $1$. This term has no affect on the spectrum or interactions
of the theory, and can be ignored.  

We
conclude that in perturbation theory, the target theory \Eq{targ} is
obtained, up to an uninteresting cosmological constant, without fine
tuning.  Given that $(a g_2)\to 0$ 
in the continuum limit, the perturbative result should be reliable,
with two caveats:  first, the continuum limit must be taken before the large
volume limit, so that the infrared $\ln N$ terms do not overwhelm
suppression by powers of $a
g_2$; and second, field values must remain small for our power
counting to be valid.  

The last point is cause for concern.  Our power counting states that an
operator of the form $\phi^2\CO$ can generate the operator $g_2^2 \CO$
at one loop by connecting the $\phi$ propagators.  Evidently, our power
counting is equivalent to assuming that fluctuations of bosonic fields
are
 roughly equal to $g_2$.
 Since there are zeromodes in the theory which
can in principle fluctuate wildly ({\it e.g.}, the radion field), we
must make sure these fields do not render our analysis invalid.

\subsection{The radion}
\label{sec:4c}
We must now deal with the zeromode $\zeta(t)$ which corresponds to the
classical flat direction of our scalar potential.  Since the Fourier
mode expansion of the link bosons is given by
$\phi_n(t) = (f + \zeta(t)){\bf 1} + \sum_{k\ne 0} e^{ik n 2 \pi/N}
\tilde \phi_k(t)$, where ${\bf 1}$ is the $k\times k$ unit matrix, it
follows that we should replace $a\to a/(1+a\zeta(t))$ and $g_2^2\to g_2^2(1+a\zeta(t))
$ through out  the Lagrangian we obtain assuming $\zeta(t)=0$.  In
addition, there will be terms proportional to time derivatives of
$\zeta(t)$.  The simplest is the kinetic term
\beq
\frac{1}{g_2^2}\int {\rm d}t\,\int{\rm d}x\, \frac{1}{2}\,\dot\zeta^2 =\frac{ L}{g_2^2}\int {\rm
  d}t\, \frac{1}{2}\,\dot\zeta^2\ .
\eeq
As there is no potential for $\zeta$, which behaves as a quantum
mechanical variable, the ground state wave function for $\zeta$ will 
be uniformly spread out among all possible values for $\zeta$, which we can take
to live on a compact space of size $>\frac{1}{a}$, if we consider the
mother theory to derive from a compactified $N=1$ SYM in 3+1 dimensions. This would ruin
our renormalization arguments of the previous section, since the
argument of the previous section implies that $\zeta$ fluctuations are
$O(g_2)$ in size; if this fails to hold, we do not have any criterion
for identifying an operator as 
irrelevant.

Note however that the kinetic term for $\zeta$ is proportional to $L$,
the size of the lattice; it acts like a heavy particle for large lattices.  It follows
that if one performs a path 
integral over $\zeta$ with localized initial ($t=0$) and final ($t=T$) wave functions
 then
$\zeta$ will remain small provided that $\Psi$ is sharply enough peaked and/or
$T$ is sufficiently small compared to $L$.  In particular, we can
consider the Euclidean time path integral 
\beq
Z=\int\,d\zeta_i \, d\zeta_f \, \Psi^*(\zeta_f)\Psi(\zeta_i)
\int\,d\zeta(t)\,e^{-\frac{L}{g_2^2}\int_0^T\,{\rm
    d}t\,\frac{1}{2}\dot\zeta^2}\ ,\qquad \zeta(0)=\zeta_i\ ,\quad \zeta(T)=\zeta_f
\eeq
where we will specify normalized Gaussian wave functions
\beq
\Psi(\zeta)=(\pi \zeta_0^2)^{-\frac{1}{4}} e^{-\frac{\zeta^2}{2\zeta_0^2}}\ .
\eeq
With this path integral, then  $\zeta$ fluctuations at $t=T/2$, for
example, can be computed to be
\beq
\frac{\bra{\Psi,T}\,\zeta(T/2)^{2n}/n!\,\ket{\Psi,0}}{\braket{\Psi,T}{\Psi,0}}=
\frac{\Gamma(\half+n)}{\Gamma(\half)\Gamma(1+n)}
\left(\zeta_0^2 + \frac{T g_2^2}{2L}\right)^n\ .
\eqn{zetafluct}
\eeq
In the above expression, the $\zeta_0^2$ corresponds to the dispersion
in $\zeta$ inherent in our initial and final conditions on the path
integral, while the $T g_2^2/2L$ term corresponds to the dispersion
resulting from the random walk of
a free particle with mass $L/g_2^2$.  
However, as we discussed in the previous section, our renormalization analysis
implicitly assumed that integrating out $\zeta$ (or any scalar field) at $\ell$ loops would involve the
 replacement in the effective action of (roughly)
$\zeta\to g_2$.
By comparing this formula with  \Eq{zetafluct} above, we see that in order to
justify our renormalization analysis of the previous section, we can
 take
\beq
\zeta_0 \to 0 \ ,\qquad T= L\ .
\eeq
By taking $\zeta_0\to 0$, we are specifying that 
the initial and final conditions on the path integral are
$\zeta(0)=\zeta(T)=0$.  (It was convenient to perform the computation
of \Eq{zetafluct} with nonzero $\zeta_0$, since both the denominator and the numerator on the
left side of the equation vanish for $\zeta_0=0$).  Since the volume $L=Na$ can be made
arbitrarily large (so long as $(a g_2)^2\ln L/a \ll 1$) our condition on
$T$ poses no obstacle to taking the continuum and infinite volume
limits.  The main point of this analysis is that the radion behaves as
a particle whose mass
 scales like the target theory's volume.  

The radion is not the only modulus of the theory;  there exist other
flat directions which generically break the $U(k)$ gauge symmetry of
the target theory down to $U(1)^k$.  The treatment of these zeromodes
is the same as  for the $U(k)$-singlet field
$\zeta$. We have shown that  it is meaningful to
talk about moduli in this particular class of quantum mechanical
theories, so long as we do not try to measure correlation functions over
times long compared to the spatial dimensions of the lattice.

These conditions mark another departure from the usual
approach to lattice field theory:  all correlation functions are
being measured in a particular excited state of the lattice theory, instead of the ground
state of the system.

We conclude that the $1+1$ dimensional  target theory with $(2,2)$ supersymmetry
\Eq{targ} may be obtained from our $0+1$ dimensional spatial lattice
theory without any fine tuning, due to the underlying exact
supersymmetry of the lattice.  It would be interesting to compare
lattice results for this theory with numerical results from the
discrete light cone approach of ref. \cite{Antonuccio:1998jg}.
It should be apparent that one can 
generalize this model to include $SU(k)$ adjoint matter fields in the target
theory, interacting via  a superpotential, by adding such $SU(kN)$
adjoint matter fields to the mother
theory.  It is also possible to add matter fields to the target theory
in the defining
representation of $SU(k)$  by adding $N$ flavors of fermions to the
mother theory, transforming as fundamentals of $SU(k N)$.  In this
case, the $Z_N$ used in the orbifold condition \Eq{orbi} would be
constructed to contain contributions from the $SU(N)$ flavor symmetry
of the mother theory.  For the case $k=1$, resulting in an Abelian
target theory, it would be interesting to
see if a lattice theory could be used to explore properties of
Calabi-Yau spaces, along the lines of \cite{Witten:1993yc}.

\section{Higher dimension  examples}
\label{sec:5}

We do not discuss here the 1+1 dimensional target theories with eight
or sixteen supercharges, all corresponding to the lattice of
Fig.~\ref{fig:lat1d} with four and eight exact supersymmetries
respectively. Instead, we give a brief discussion of the higher
dimension 
theories corresponding to the lattices pictured in Figs.~3-5. A  more
detailed analysis is in preparation.

\subsection{Eight supercharges in 2+1 dimensions}
\label{sec:5a}

The two dimensional lattice of Fig.~\ref{fig:lat2dsquare}  again respects  two exact
supercharges.  The target theory, equivalent to $N=2$ SYM reduced from
$3+1$ to $2+1$ dimensions, has been discussed in
refs.~\cite{Seiberg:1996bs,Seiberg:1999xz}.  It is a gauge theory with
two complex two-component adjoint fermions, three real adjoint scalar fields, and a
gauge field; all interactions are related to the gauge coupling $g_3$.
 Fig.~\ref{fig:EJ2dsquare} displays a single cell of our two
 dimensional lattice.  The horizontal and vertical 
links are bosonic chiral superfields $\Phi_{x,y}$, while the diagonal links are
Fermi superfields $X$. As discussed in the appendix,  Fermi
superfields $ X_a$ in general satisfy $\mybar\CD
 X_a = E_a(\Phi)$, where $E_a(\Phi)$ is a holomorphic chiral superpotential
 satisfying $\mybar \CD E_a(\Phi)=0$.  Then besides the kinetic parts
 of the Lagrangian for the chiral, Fermi, and vector supermultiplets
 ($L_\Phi$, $L_X$ and $L_g$ respectively), the $X_a$  can be
coupled to a second holomorphic chiral superpotential $J^a(\Phi)$ satisfying
$E_a J^a=0$, through a term in the Lagrangian of the form
$L_J=\frac{1}{\sqrt{2} g^2} X_a J^a\vert_{\theta}$. Note that $E$ must be in the same
representation of the gauge group as $X$, while $J$ must be in the
conjugate representation.

 We have drawn our lattice indicating 
chirality arrows for our chiral (dark blue) and Fermi (light blue)
link superfields; a closed path
on the lattice that consists of a single Fermi superfield link, and
any number of oriented (head to tail) chiral superfield links
constitutes an $E$ or a $J$ interaction---if the Fermi link is
oriented in the same sense along the path as all of the chiral
superfield links, then the plaquette term appears in the Lagrangian as
a $J$ interaction; if the arrow of the Fermi link runs counter to the
chiral superfield links in the plaquette, then that plaquette
interaction arises from the $E$ superpotential.  Fig.~\ref{fig:EJ2dsquare} makes it
evident that only $J$-type interactions occur in this theory, so that
$\mybar\CD X_a=0$ for all of the Fermi supermultiplets. Each
Fermi link borders on two triangular plaquettes of opposite
orientation; they are assigned a relative  sign.
 Since the mother theory has no interactions
higher than cubic in the superfields, we need only consider these three-link
plaquettes as shown. 

  We label each site by the two-component
vector ${\bf n}=\{n_x,n_y\}$;   $\Phi_x({\bf  n})$ and $\Phi_y({\bf
  n})$ denote the  horizontal and vertical  link variables (chiral
superfields) pointing
outward from the site ${\bf n}$ ({\it e.g.} in the $+{\bf \hat x}$ and
$+{\bf \hat y}$ directions respectively); $X({\bf n})$ is the Fermi superfield on the
diagonal link pointing toward the site ${\bf n}$ (in the $-({\bf \hat
  x}+{\bf \hat
y})$ direction).
  Then $L_J$ for this theory is
given by
\beq
L_{J} = \frac{1}{g^2}\sum_{\bf n}\,  \Tr \left[X({\bf n}) \Bigl(\Phi_x({\bf
n})\Phi_y({\bf  n + \hat x}) - \Phi_y({\bf  n})\Phi_x({\bf  n + \hat
y})\Bigr)\right]{\Big\vert_{\theta}} + h.c.\ .
\eqn{sup}\eeq

 Similar to the previous example of \S\ref{sec:4},
we choose to expand fields about the point in the classical moduli space
$\phi_x({\bf n})=\phi_y({\bf n})=f/\sqrt{2}$ times the $k\times k$ unit
matrix. It is important for our renormalization arguments that we have
chosen a point in the classical moduli space that preserves the $C_{2v}$
lattice symmetry. The continuum limit is taken by taking $f=1/a$, $a\to 0$,
keeping fixed $g^2_3\equiv a^2 g^2$.  In the continuum limit,
the above interaction \Eq{sup} yields the transverse kinetic term for
the gauge bosons, $F_{xy}^2$,   as well  as parts of the kinetic terms
for the scalars and fermions.

\DOUBLEFIGURE[t]{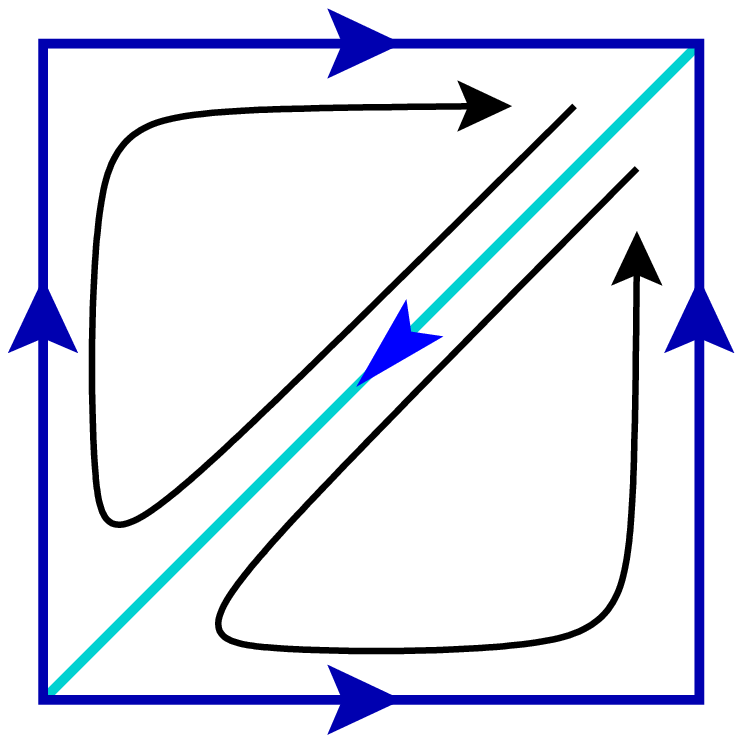,height=4cm}{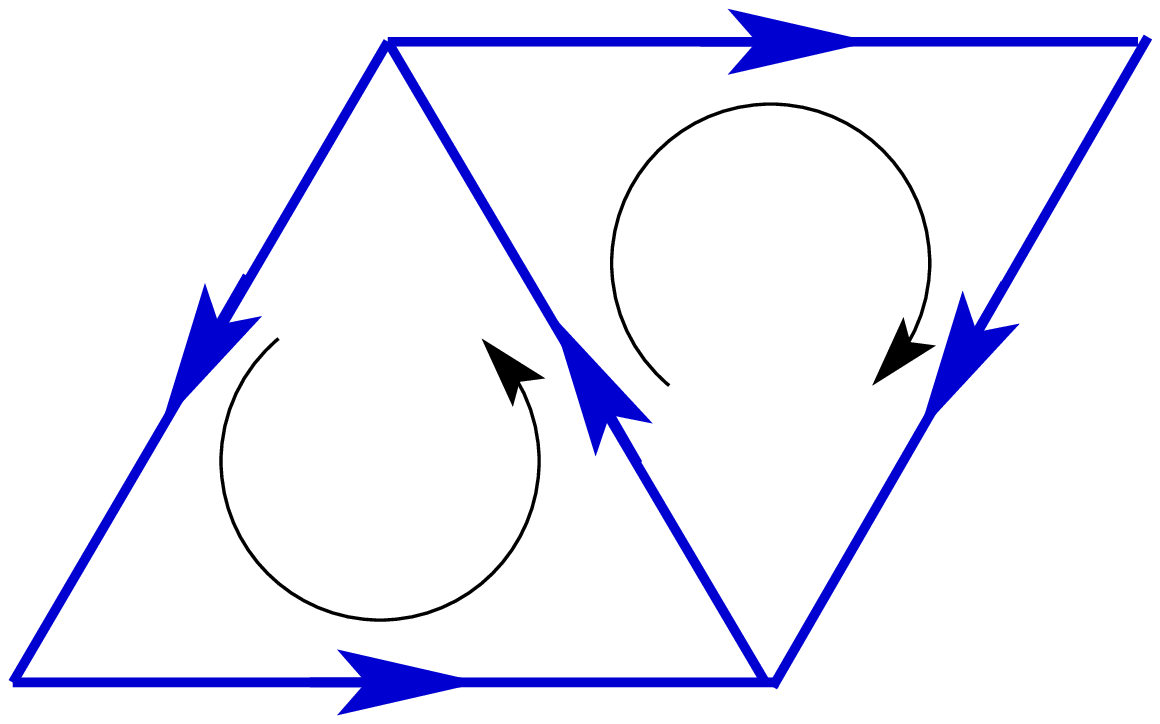,height=4cm}
{\sl One cell from the lattice of Fig.~3.  The black arrows indicate the plaquette terms in
  $L_J$, with opposite orientations contributing with opposite signs.
\label{fig:EJ2dsquare}}
{\sl Two plaquettes contributing with opposite sign to the
  superpotential $W$ in the lattice  for the target theory of sixteen
  supercharge SYM  in $2+1$ dimensions (Fig.~5).
\label{fig:EJ2dhex}}%

Analysis of the continuum limit for this theory is similar in spirit to the
analysis of \S\ref{sec:4c}.  Loops introduce powers of $(a g_3^2
)^{\ell}$, and so in principle operators with dimension $p\le
(4-\ell)$ may be renormalized at $\ell$ loops, where $p$ is given in
Table~2. With the exception of the cosmological constant, which can
arise as a Fayet-Iliopoulos term, none of the operators in that table with $p\le 3$  are
consistent with the underlying supersymmetry and crystal symmetry of
the lattice.  For example, under reflections about the diagonal axis,
the have $\Phi_x(\bfn)\leftrightarrow\Phi_y(\tilde\bfn)$ and $X(\bfn)\to
-X(\tilde\bfn)$, where $\tilde\bfn=\{n_y,n_x\}$.  This symmetry (along with the exact supersymmetry and
gauge symmetry)
precludes us from obtaining  $\phi$, $\phi^2$, 
$\phi^3$ or $\phi\nabla \phi$  terms in the scalar potential. 
The $p=3$ operators $\phi^*\partial_0\phi$ and $\mybar\psi\psi$ are
allowed by both the crystal symmetry and supersymmetry, arising from
the operator
$\bigl.\mybar\Phi(\bfn)\Phi(\bfn)\bigl\vert_{\mybar\theta\theta}$, but
are forbidden by time reversal symmetry. Therefore no fine tuning of
the tree level theory is required to obtain the desired target theory
in the continuum limit.

As in the $1+1$ dimensional case, the lattice spacing in the $2+1$
dimensional theory is dynamical and
we have a radion --- in fact there is one each for the horizontal and
vertical link directions.  In
this case the inertia of each radion scales as $N^2$, the volume of the
target theory.   The treatment of the radion 
is as in the lower dimension theory, \S\ref{sec:4c}.

\subsection{Sixteen supercharges in 2+1 dimensions}
\label{sec:5b}

The action corresponding to the hexagonal lattice of Fig.~\ref{fig:lat2dhex} possesses
an exact $D_6$ symmetry and four exact supercharges.  The notation for
this theory is similar to the familiar $N=1$ supersymmetry in $3+1$
dimensions, which also has four supercharges.  The three link
variables emanating from each site are
described by chiral superfields $\Phi_i$, $i=1,2,3$, each containing 
a complex scalar and two complex fermions, which can be thought of as
the dimensionally reduced chiral superfield from an $N=1$,
$3+1$-dimensional theory.
Vector superfields $V$ reside at each site, containing a
real gauge field $v_0$, three real scalars, and two complex fermions
--- the dimensional reduction of a four dimensional gauge field and
a Weyl fermion gaugino.  There exists a superpotential $W$ which
contains the triangular plaquette terms, signed according to
orientation.  We can define three lattice vectors
\beq
{\bfm_1} = (1,0)\ ,\quad
\bfm_2=\left(-\frac{1}{2},\frac{\sqrt{3}}{2}\right)\ , \quad
\bfm_3=\left(-\frac{1}{2},-\frac{\sqrt{3}}{2}\right)\ , 
\eeq
and by $\Phi_i(\bfn)$ denote the chiral superfields leaving site $\bfn$ along the
$\bfm_i$ direction. We can write the superpotential $W$ in terms of
these fields as
\beq
W = \sum_{\bf n} \,
\Tr\Phi_2(\bfn)\left(\Phi_3(\bfn+\bfm_2)\Phi_1(\bfn-\bfm_1) -
  \Phi_1(\bfn+\bfm_2)\Phi_3(\bfn-\bfm_3) \right)
\eeq
corresponding to plaquettes of Fig.~\ref{fig:EJ2dhex}.

We expand this theory about $\phi_i=\frac{f}{\sqrt{2}}$ times the $k\times k$ unit
matrix, where  now $f$ is
related to the lattice spacing $a$ as $f=\sqrt{3/2}/a$. The gauge
coupling is defined again as $g_3^2\equiv a^2 g^2$, and it is
held fixed in the $a\to 0$ and $N\to\infty$ limits. As in the previous
example of a target theory in $2+1$ dimensions with 
eight supercharges, there are no $p\le 3$ operators from Table~2 that can be induced as
counterterms in the  continuum limit, other than a cosmological
constant.  Therefore one may expect to
obtain the desired supersymmetric target theory without any fine
tuning. 

\subsection{Sixteen supercharges in 3+1 dimensions: the BCC lattice}
\label{sec:5c}

The three dimensional lattice of Fig.~\ref{fig:lat3dBCC} is our third example
respecting
 two exact supercharges.  This model is of particular interest because
the target theory is $N=4$ SYM with a $U(k)$ gauge group
in 3+1 dimensions.  This target theory  can be written
in $N=1$ notation in terms of three $k\times k$ chiral superfields $\Phi_i$
transforming as $\Phi_i\to U^\dagger \Phi_i U$ under the $U(k)$ gauge
symmetry.  These fields interact via a
superpotential $W=i\epsilon_{ijk}\Tr \Phi_i \Phi_j\Phi_k$, as well as through gauge interactions.

Similar to the 2+1 dimensional example
discussed above, the lattice theory consists of vector, chiral and Fermi
supermultiplets. A new feature of the $3+1$ dimensional lattice is
that there exist both $E$- and $J$-type plaquette terms.  Each Fermi
multiplet forms one
edge of four triangular plaquettes, as shown in
Fig.~\ref{fig:EJ3dBCC}.  Two of the four plaquettes run in the direction 
of the Fermi
multiplet orientation, and are therefore of the $J$-type discussed
previously; the other two run in a sense counter to the Fermi link's
orientation and form the $E$-type interactions.  Each $E$- and $J$-type
plaquette comes in two opposite orientations, and therefore the two contributions of
each type will have a relative minus sign.  The discrete symmetries of
the lattice ($e.g.$ the $C_4$ symmetry evident in Fig.~\ref{fig:EJ3dBCC} ensure that
the $E$ and $J$ superpotentials have the same 
form and strength. All superpotentials we write down have the form of
a product of two chiral superfields and one Fermi superfield,
corresponding to the three edges of the plaquette.

\EPSFIGURE[t]{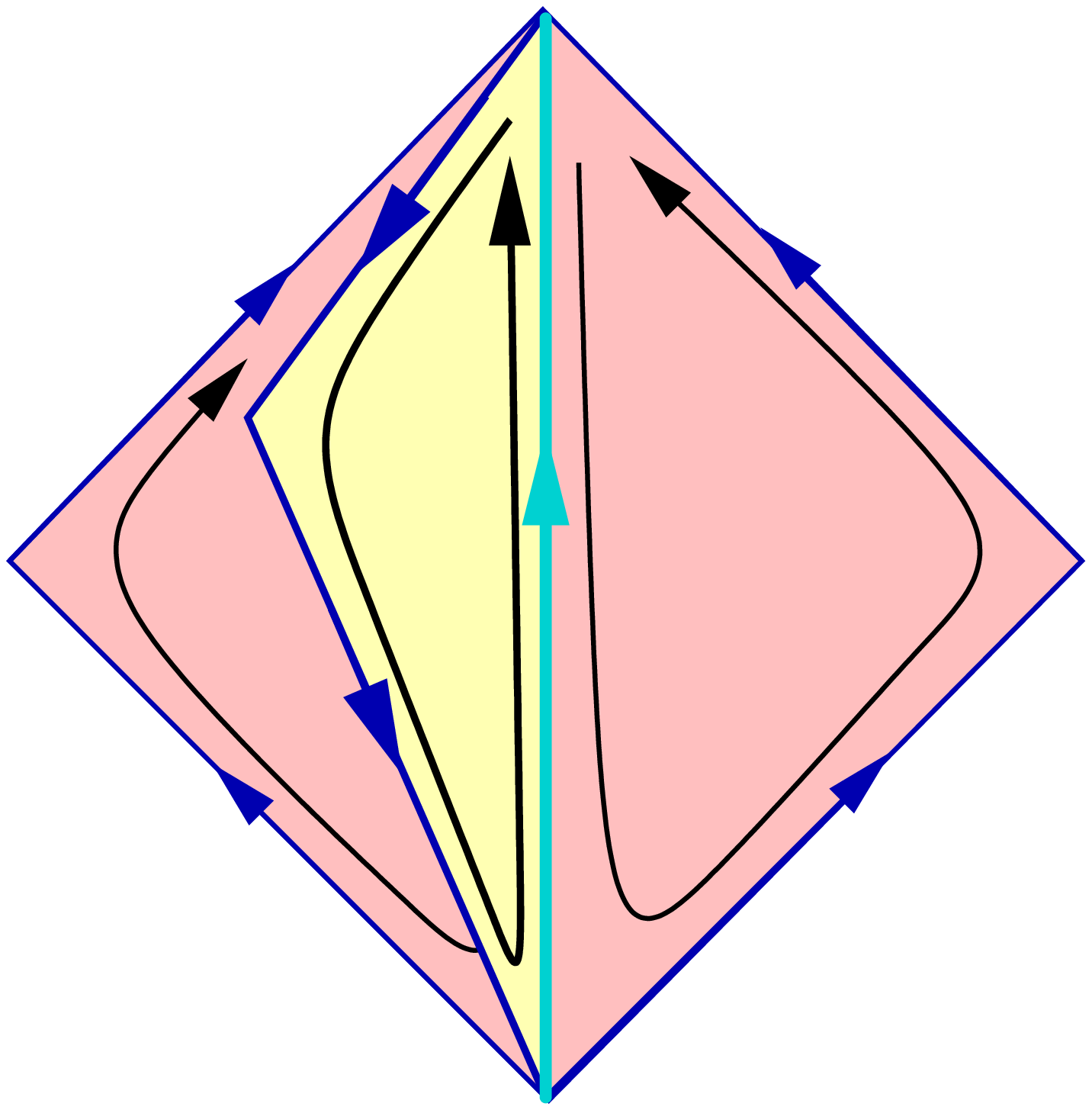,height=6cm}
{\sl Each Fermi supermultiplet (light blue link) in the body centered
  cubic crystal of 
  Fig.~7 forms the edge of four triangular plaquettes: one of each orientation of
  both the $E$  (pink) and $J$ (yellow) types.  
\label{fig:EJ3dBCC}}%


To write down the Lagrangian, we denote each lattice site in the
body centered cubic crystal by a three dimensional dimensionless vector ${\bf n}$,
where we take the edge of the cell in Fig.~\ref{fig:lat3dBCC} to be of coordinate length 1.
Each site is associated with three Fermi multiplet links pointing
outward in
the ${\bf \hat x}$,  ${\bf \hat y}$, and  ${\bf \hat z}$ directions,
and are denoted by $X_a({\bf n})$ respectively, where $a=1,2,3$.
Each site is also associated with  four bosonic chiral superfield
links, $\Phi_\mu$, $\mu=0,\ldots,3$, oriented outward from the site in the four directions
\beq
{\bf m}_0 = \left(\half,\half,\half\right) \ ,\quad 
{\bf m}_1 = \left(\half,-\half,-\half\right) \ ,\quad 
{\bf m}_2 = \left(-\half,\half,-\half\right)\ ,\quad 
{\bf m}_3 =\left(-\half,-\half,\half\right)  
 \ .
\eqn{meq}
\eeq
Then for each site $\bfn$, we construct the six
superpotentials $E_a$ and $J_a$, each with two terms of opposite sign,
corresponding to the twelve triangular plaquettes associated with that
site: 
\beqa
E_a(\bfn)&=&\sqrt{2}\,\delta_{ja}\left(\Phi_0(\bfn)\Phi_j(\bfn+\bfm_0)-\Phi_j(\bfn)\Phi_0
  (\bfn+\bfm_j)\right)\
,\cr &&\cr
J^a(\bfn)&=&\sqrt{2}\,\epsilon_{ija}\left(\Phi_i(\bfn+{\bf \hat r_a})\Phi_j(\bfn-\bfm_j)\right)\
.
\eqn{ej}
\eeqa
where ${\bf \hat r_a}$ equals $\xh$, $\yh$ or $\zh$ for
$a=1,2,3$. For example, defining $\zh$ to point in the vertical
direction, and $\bfm_3$  and $\bfm_0$ lie in the plane of this page, the two
$\Phi$ bilinears in $E_3$ correspond to the dark pink triangular
plaquettes in Fig.~\ref{fig:EJ3dBCC}, while the two $\Phi$ bilinears in  $J^3$
correspond to the two light yellow plaquettes, one of which is hidden
in the figure.
It is not hard to show from \Eq{ej} that $\Tr E_a J^a=0$ as required
by supersymmetry, by considering 
all contributions to any particular ordering of the $\Phi$ fields
($e.g.$, $\Tr\Phi_0\Phi_1\Phi_2\Phi_3$), making use of the
invariance of the lattice under translation by the $\bfm$ vectors and identities such as
$\bfm_2+\bfm_3=-\xh$.

To create the necessary hopping terms to generate the extra
dimensions, we again choose a point in the classical moduli space which
preserves the $O_h$ lattice symmetry, namely $\Phi_\mu = f/\sqrt{2}$ times
the $k\times k$ unit matrix, where $f=1/a$,  $a$ being the lattice spacing.
  As before, there will be radion
zeromodes associated with fluctuations of the lattice spacings, which
must be fixed in the path integral as in \S\ref{sec:4c}. The
continuum limit involves taking $a\to 0$ keeping the four dimensional
gauge coupling $(g_4)^2 = a^3 g^2$ fixed. At tree level one can show
that indeed one obtains the target 
theory of $N=4$ SYM in $3+1$ dimensions. Defining the matrix
$m_{\mu\nu}$, where $\mu,\,\nu=0,\ldots,3$ as
\beq
m_{\mu\nu} = \frac{1}{2}\begin{pmatrix}
\ \ 1 &\ \ 1&\ \ 1&\ \ 1\cr
\ \ 1 & \ \ 1& -1 & -1\cr
\ \ 1&-1& \ \ 1&-1\cr
\ \ 1&-1&-1&\ \ 1
\end{pmatrix}
\eqn{mvecs}\eeq
we can express the three gauge fields $v_{1\ldots3}$ and six real scalar
fields $s_{1\ldots 6}$ of the continuum $N=4$ SYM theory as
 \beq
\frac{s_i+iv_i}{\sqrt{2}}= m_{i\mu}\phi_\mu\ ,\qquad
\frac{s_4+is_5}{\sqrt{2}}=  m_{0\mu}\phi_\mu\ ,\qquad
s_6=\sigma\ .
\eeq
The $F_{ij}^2$ kinetic terms for the $v_i$ arise from
$V_E+V_J$; the $F_{0i}^2$ kinetic terms come from the
$|D_0\phi_\mu|^2$ operators in $L_\Phi$. The
spatial parts of the
kinetic terms for the $s_{1\ldots 5}$ scalars arise from $V_d$ and $(V_E+V_J)$, with unwanted Lorentz
violating contributions canceling between the two. The field
$s_6$ develops a hopping term from the $|[\sigma,\phi]|^2$ operator in
$L_\Phi$.  In a similar manner it is possible to construct the four
Weyl fermions of the continuum theory from the eight single component
fields $\lambda$, $\chi_i$, and $\psi_\mu$. There are no unwanted
doublers in our formulation.

Renormalization in this theory is trickier to analyze than in the
super-renormalizable examples discussed above.  Loops bring in powers
of $g_4^{2\ell}$, and so all operators in Table~2 with $p\le 4$ are
susceptible to infinite renormalization. However, for the same reasons
discussed in sections \ref{sec:5a}, \ref{sec:5b} for the $2+1$ dimensional theories with
eight and sixteen supercharges, none of the dimension $0<p\le 3$
counterterms of Table~2 are allowable by the exact lattice symmetries.  Therefore we
need only consider the perturbatively marginal  $p=4$ operators. To
classify these $p=4$ operators which are consistent with both the
underlying $O_h$ and supersymmetries is somewhat complicated; however
it is apparent that there are now a number of allowed operators not
included at tree level, which could potentially involve logarithmic fine tuning to
attain the desired target theory.  We have not ascertained whether fine
tuning is actually necessary in this model, but will instead offer a
variant of this theory which looks promising to being stable against renormalization.

\subsection{Sixteen supercharges in 3+1 dimensions: the asymmetric lattice}
\label{sec:5d}

\DOUBLEFIGURE[t]{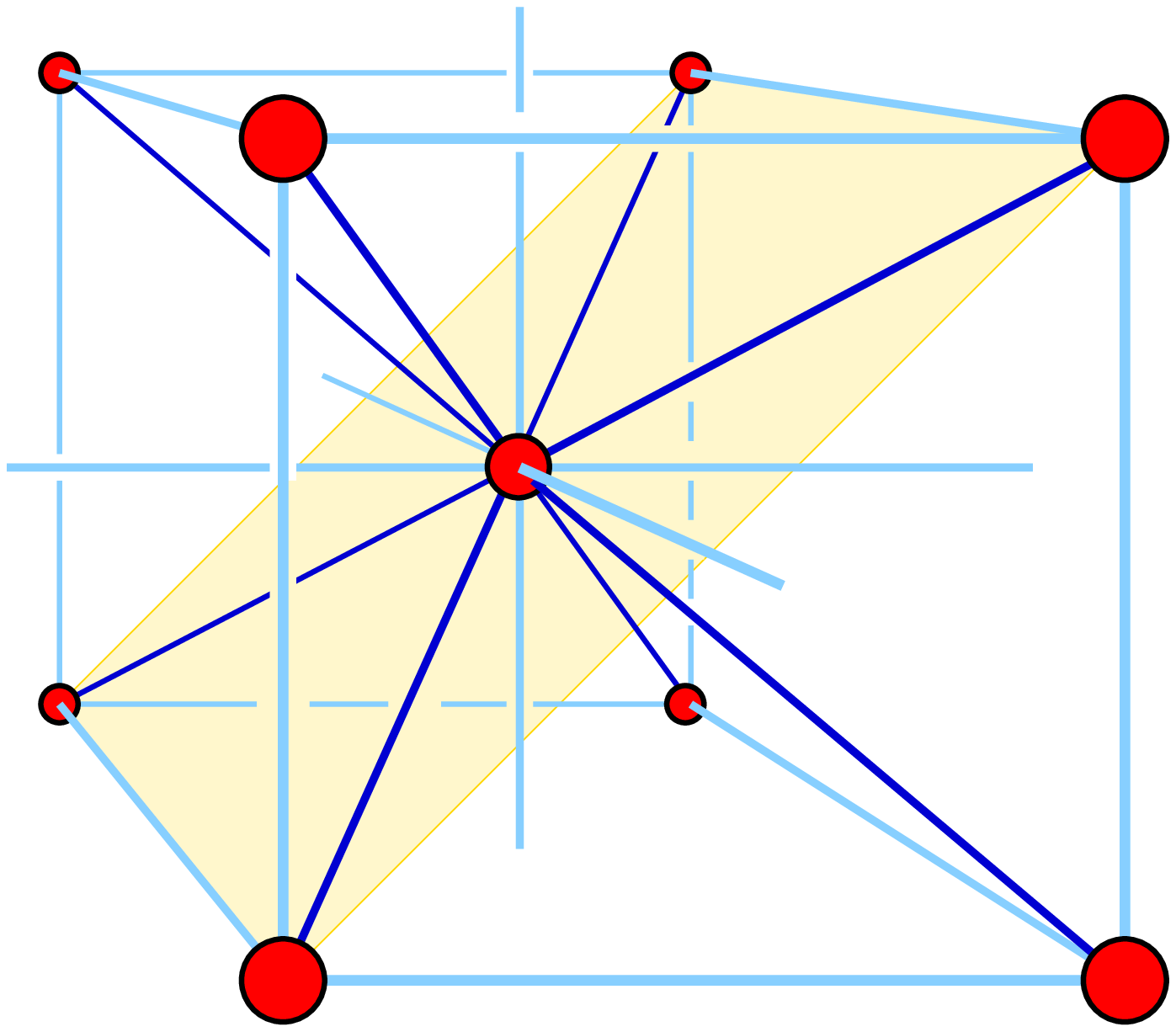,width=5cm}{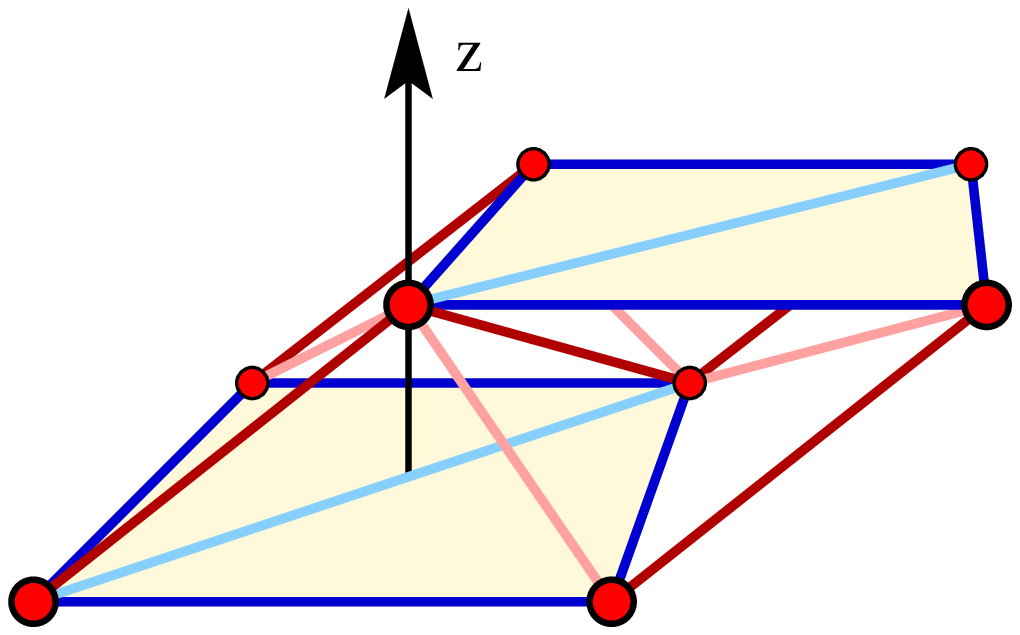,width=7cm}
{\sl The two dimensional sublattice of the bcc lattice is essentially
  the same as the one encountered in Fig.~3, and if isolated leads to
  a $2+1$ dimensional continuum theory with eight supercharges,
  without fine tuning (see \S5.1).
\label{fig:asymmBCC}}
{\sl A redrawing of the three dimensional lattice for $N=4$ SYM in 3+1
  dimensions.  Dark blue and dark red links
  correspond to chiral supermultiplets; light blue and pink links are
  Fermi supermultiplet.  The continuum limit is taken
  asymmetrically, 
  in the horizontal planes first and subsequently in the $z$
  direction. After the first limit is taken, the theory is the
  ``linear moose''  shown in
  Fig.~13.\label{fig:asymmrhombus}} %

It may be  profitable to consider an asymmetric version of the BCC lattice of
Fig.~\ref{fig:lat3dBCC}.  Instead of choosing to take the continuum limit uniformly in
each link direction, we choose instead to take the continuum limit in
two dimensional planes of the lattice, and then subsequently take the
continuum limit in the direction orthogonal to these planes. Note
that the two dimensional sublattice lattice in these planes is
essentially the same as was discussed in \S\ref{sec:5a}, as seen in
Figs.~\ref{fig:asymmBCC},~\ref{fig:asymmrhombus}. The
advantage in first taking the continuum limit in these planes is that
the first step produces a family of continuum $2+1$ 
dimensional gauge theories with bifundamental matter fields, for which
the fine tuning problems will be less serious than in $3+1$ dimensions, due to the
super-renormalizability of the target theory.  The
subsequent continuum limit in the third spatial direction then benefits by
the existence of six additional conserved supercharges which have emerged in
the $2+1$ dimensional theory; these supercharges can help  protect against
the logarithmic fine tuning associated with dimension four operators.

\EPSFIGURE[t]{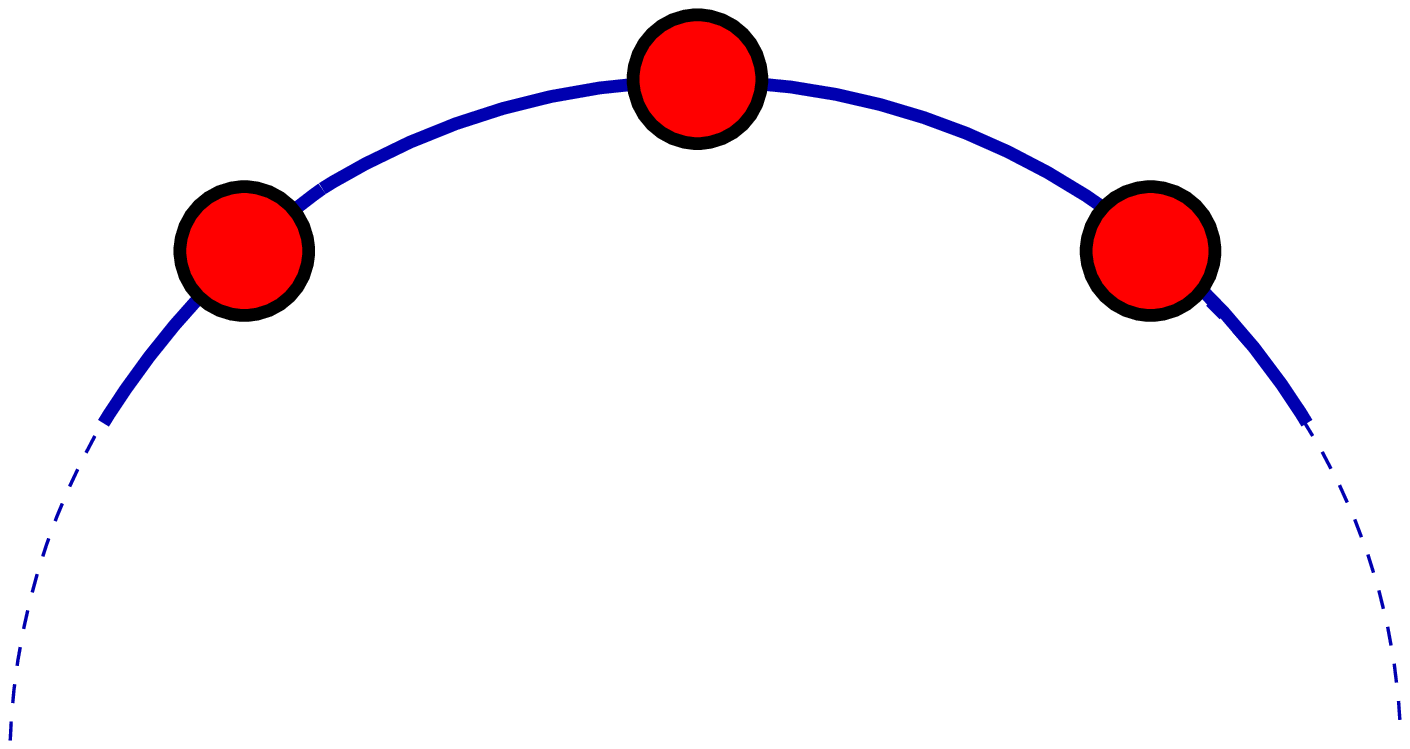,width=7.1cm}
{\sl A deconstruction of $N=4$ SYM theory in terms of a linear
  ``moose'' in $2+1$ dimensions with eight exact supersymmetries.
  This is the dimensional reduction of an $N=2$ SYM theory in $3+1$
  dimensions with a bifundamental hypermultiplet at each link. This is
  the desired intermediate target theory in for the lattice in
  Fig.~\ref{fig:asymmrhombus} after the continuum limit has been taken in the horizontal
  planes.
\label{fig:linearmoose}}%
To be more specific, we propose to define the continuum limit by setting the link vevs in
the horizontal plane (blue, in Fig.~\ref{fig:asymmrhombus}) to 
$\frac{f}{\sqrt{2}}$, while the links connecting the planes (red, in
Fig.~\ref{fig:asymmrhombus}) take the value $\frac{f'}{\sqrt{2}}$.
The lattice spacings are $a_\perp = 1/f$, $a_\parallel =1/f'$.  We first
take the continuum limit in the horizontal planes, $a_\perp\to 0$,
$N\to \infty$, keeping $g_3^2=g^2 a_\perp^2$, $N'$ and $a_\parallel$ fixed.  The desired
target theory at this stage is 
 deconstruction \cite{Arkani-Hamed:2001ca} of
$N=4$ SYM in $3+1$ dimensions in terms of a moose in $2+1$ dimensions
where at each site we have a gauge theory with eight supercharges, and
on each link we have a hypermultiplet, shown in Fig.~\ref{fig:linearmoose}.  This hypermultiplet (we use
the $3+1$ dimensional, $N=2$ language) consists the variables living
on the links connecting the planes (red) in
Fig.~\ref{fig:asymmrhombus}: four complex fermionic and two complex
bosonic degrees of 
freedom. The second stage of our continuum limit involves taking the
lattice spacing $1/f'=a_\parallel\to 0$ in the $2+1$ dimensional
moose.  

To understand the number of parameters that need to be fine tuned in
order to obtain the desired $N=4$ continuum theory from our proposed
lattice, one can look at the two continuum limits separately.  At the
first stage we expect the lattice symmetries to forbid counterterms
for dimension $p\le 3$ operators, in a similar manner as they did for
the simpler example of a $2+1$ dimensional target theory with eight
supercharges discussed in \ref{sec:5a}. When we subsequently take the
$a_\parallel\to 0$ limit,  there is one relevant parameter to consider
in the moose of Fig.~\ref{fig:linearmoose}, namely a mass term for the hypermultiplets. It
seems plausible that in fact that the conformal fixed point
corresponding to $N=4$ SYM in $3+1$ dimensions occurs at zero ``bare'' hypermultiplet mass in
the moose, which would obviate fine tuning at this second stage.

We leave open the number of parameters which need to be fine tuned to
obtain the $N=4$ SYM theory in $3+1$ dimensions, but we believe there
is reason to be optimistic that there need be no fine tuning at all
following the prescription outlined here.

\section{Discussion}
\label{sec:6}
Motivated by deconstruction and utilizing the orbifolding techniques
developed for string theory, we have shown a method for constructing
supersymmetric Yang-Mills theories with extended supersymmetries  on
spatial lattices of
various dimensions.  Several of these theories ---
in particular, the 
ones with sixteen supercharges --- are expected to exhibit interesting
nontrivial conformal fixed points in the infrared. The basic approach
relies on maintaining exactly a subset of the supersymmetries desired
in the continuum limit.   It is remarkable that this allows one to
describe interacting scalars on the lattice without any fine tuning of
parameters in the continuum limit. In fact we have argued that none of the
four, eight or sixteen supercharge target theories in $1+1$ or $2+1$
dimensions require any fine tuning.  

We have also given a prescription for latticizing the extremely interesting case of $N=4$ SYM
in $3+1$ dimensions.  We are optimistic that our proposal can be used
to study this theory without any fine tuning of parameters, but this
remains an open question. In any case, the degree of
tuning is expected to be substantially less than by using
conventional latticization approaches.

While we have
focused on pure Yang-Mills theories, it is possible to generalize
these theories to include matter fields as well for the target theories with four
or eight supercharges.  For example, in the lattice model discussed in detail
for (2,2) SYM in $1+1$ dimensions, matter can be incorporated by
including $N$ flavors of chiral superfields at each lattice site, and
by arranging the  $Z_N$ symmetry of the orbifold to reside in part
within  the $SU(N)$ flavor symmetry of the matter fields.  

The theories on spatial lattices are only of limited use for numerical
investigation, although one can imagine combining strong coupling
expansions and large-$N$ expansions (in our notation, large-$k$
expansions) to extract information about the target theories.  It is
known, for example that in the large $k$ limit, the planar diagrams of the
daughter theories with rescaled coupling exactly agree with those of
the mother theory.  For  actual
numerical simulation, the techniques described here can be
extended to Euclidean space-time lattices.
From Table~1 one sees that by reducing SYM 
theories down to zero dimensions, in each case the rank of the $R$
symmetry group $G_R$ increases by one.  That means that all of the
target theories discussed here can in principle be realized from pure
spacetime lattices possessing half of the exact supersymmetry of the
analogous spatial lattices.  For example the theories with sixteen
supercharges in $2+1$ and $3+1$ dimensions can be constructed from
spacetime lattices possessing two or one exact real supercharges
respectively.   A series of papers describing such a construction are
in preparation, the first of which is Ref.~\cite{Cohen:2003xe}.

An open question of interest is whether one can construct lattices
with enough exact supersymmetry to force the emergence of supergravity
in the continuum limit.

\acknowledgments
We would like to thank Andrew Cohen, Andreas Karch and Guy Moore for
useful conversations.  D.B.K. and M.U. were supported in part by DOE
grant DE-FGO3-00ER41132, and E.K by DOE grant DE-FG03-96ER40956.

\appendix
\section{Superfield notation for 2 supercharges in 0+1 dimensions}
\label{sec:apnotation}

In this appendix we summarize the superfield notation needed for this
paper for SUSY quantum
mechanics with two real supercharges. We follow the notation of
ref. \cite{Witten:1993yc} (see also \cite{Garcia-Compean:1998kh}) for
$(0,2)$ supersymmetry in $1+1$ dimensions, which we dimensionally
reduce to $0+1$ dimensions. Superspace is parametrized by a single
complex Grassmann coordinate $\theta$ and its complex conjugate,
$\mybar\theta$. Superfields are functions of $t$, $\theta$ and
$\mybar\theta$.  The complex supercharge $Q$ and spinor derivative $D$ are given by
\beq
Q = \partial_\theta + i \bar\theta\partial_t\ ,\qquad D =
\partial_\theta - i \bar\theta\partial_t\ .
\eeq
The fields we will consider are the vector, chiral, and Fermi
superfield.  In 
Wess-Zumino gauge the vector field $V$ consists of a gauge field
$v_0$, a real scalar field $\sigma $, and a complex, one-component
fermion $\chi$, and an auxiliary field $d$ (all in the adjoint
representation)\footnote{The field $\sigma $ descends from the $v_1$
  gauge field in $(0,2)$ 
supersymmetry in $1+1$ dimensions; however it does not play the role
of the $v_1$ field in our lattice models, so we have renamed it.}:
\beq
V=(v_0 -\sigma   ) -2i\theta
\mybar \lambda  -2i \mybar \theta \lambda  - 2\mybar\theta\theta d 
\ .
\eqn{vdef}\eeq 
Chiral superfields $\tilde\Phi$ satisfy $\mybar D\tilde\Phi=0$ and have the
expansion 
\beq
\tilde\Phi =\phi + \sqrt{2}\,\theta\psi+i\mybar\theta\theta \dot \phi
\eqn{bchi1}
\eeq
where $\phi$ and $\psi$ are complex boson and Grassmann fields
respectively.  In order to implement gauge invariance, we introduce
the a gauge chiral superfield $\Psi$, which in Wess-Zumino gauge is
\beq
\Psi = -\mybar\theta \theta(v_0  + \sigma   )
\eeq
and we define the gauge covariant supersymmetric and ordinary  derivatives
\beq
\CD^{+} = e^{-\Psi}De^\Psi\ ,\qquad \CD_0^{-} = \partial_t + i V\
,\qquad D_0 = \partial_t + 
i v_0\ ,\qquad D_0^{\pm} = \partial_t +
i (v_0\pm\sigma )\ .
\eeq

It is convenient to define a chiral superfield $\Phi$ (no tilde)
which satisfies a gauge covariant chirality condition,
\beq
\Phi \equiv  e^{\Psi}\tilde\Phi = \phi + \sqrt{2}\,\theta\psi+ i\mybar\theta\theta
D^+_0\phi 
\ , \qquad \mybar\CD \Phi=0\ .
\eeq

  The
gauge field strength is contained in the
superfield $\Upsilon$ defined by
\beq
\Upsilon = [\mybar\CD^+,\CD^-_0]= -2\lambda + 2\theta(id + D_0\,\sigma  )
-2i\mybar\theta\theta D^+_0\,  \lambda\ .
\eqn{upsdef}\eeq

In addition, we must consider the so-called Fermi multiplets, Grassman
superfields $X$ satisfying $\mybar\CD X = E(\Phi)$, where $E$ is
a holomorphic function of chiral superfields.  The new components of
$X$ consist of a complex fermion $\chi$ and an auxiliary
field $G$:
\beq
X = \chi -\sqrt{2}\,\theta G+i\mybar\theta\theta D^{+}_0 \chi
-\sqrt{2}\,\mybar\theta E(\Phi) 
\eqn{fermi}
\eeq

 We normalize our
fields so as to bring a common factor of $1/g^2$ in front of the
action; all fields have canonical normalization for $g=1$.  The
Lagrangian for superfields $\Phi_j$ and $X_\alpha$ (where $j$ and
$\alpha$ signify flavors) is then comprised of several terms\footnote{The adjoint
  fields are written as matrices, $v_0=v_{0,a}T^a$, $\sigma=\sigma_a
    T^a$, where $T^a$ are the generators of the gauge group.  In $L_g$
    the $T^a$ are in the fundamental representation, normalized as
  $\Tr T_a T_b=\delta_{ab}$. In $L_\Phi$ and $L_X$, the
  $T^a$ generators are in the representation  of the $\Phi$ and
  $X$ fields.}:
\beqa
L_g &=&  \frac{1}{8 g^2}
  \Tr \mybar\Upsilon\Upsilon\Bigr\vert_{\mybar\theta\theta} \cr
&=& 
\frac{1}{ g^2}\Tr\left(\frac{1}{2}(D_0\, \sigma  )^2 + \,
   \mybar{\lambda}\,iD^+_0 \,\lambda + \frac{1}{2}d^2\right)\ ,\cr &&\cr
L_\Phi &=& \frac{i}{2 g^2}\sum_j \left(
\mybar\Phi_j\CD^-_0 \Phi_j \right)_{\mybar\theta\theta}\cr
&=&
\frac{1}{ g^2} \sum_j\left(|D_0\phi_j|^2 - \left| \sigma  \phi_j\right|^2 +
  i\mybar\psi_j D_0^-\psi_j + \mybar\phi_j d \phi_j -
  \sqrt{2}(i\mybar\phi_j\lambda \psi_j + h.c.)\right)\cr &&\cr
L_X &=& \frac{1}{2 g^2}\sum_\alpha
  \left( \mybar X^\alpha
  X_\alpha\right)_{\mybar\theta\theta}\cr 
&=&
\frac{1}{ g^2}\sum_\alpha \left( i\mybar\chi_\alpha D_0^+\chi_\alpha +
  |G_\alpha|^2 -|E_\alpha(\phi)|^2 
  -\sum_j\left[\mybar \chi^\alpha\frac{\partial E_\alpha}{\partial \phi_j}
    \psi_j +h.c.\right]\right) 
\eqn{lopsi}\eeqa
In the above expressions
$(\ldots)_{\mybar\theta\theta}$ signifies taking the coefficient of
$\mybar\theta\theta$ from the superfield in the parentheses. Further
interactions can be included through a second holomorphic 
potential $J^\alpha (\Phi)$ which satisfies $\sum_\alpha E_\alpha J^\alpha =0$:
\beq
L_J = \frac{1}{ \sqrt{2}\,g^2} \sum_\alpha\left( X_\alpha
    J^\alpha(\Phi) \right)_\theta + h.c.
=
- \frac{1}{ g^2}\sum_\alpha \left( G_\alpha J^\alpha(\phi)  + \sum_j\chi_\alpha\frac{\partial
    J^\alpha(\phi) }{\partial\phi_j}\psi_j \right)+h.c.
\eqn{lopsii}
\eeq

Both $d$ and $G_\alpha$ are auxiliary fields; integrating them out yields
the scalar potential:
\beq
V(\phi_j) = V_d + V_G + V_E = 
\frac{1}{2g^2}\sum_a\left(\sum_j \mybar\phi_j T_a
\phi_j\right)^2 +\frac{1}{g^2}\sum_\alpha\left(
\left|E_\alpha(\phi)\right|^2 + \left|J^\alpha (\phi)\right|^2\right)
\eeq

\bibliography{latticeSUSY}
\bibliographystyle{JHEP} 
\end{document}